
\magnification=\magstep1\hsize=13cm\vsize=20cm\overfullrule 0pt
\baselineskip=13pt plus1pt minus1pt
\lineskip=3.5pt plus1pt minus1pt
\lineskiplimit=3.5pt
\parskip=4pt plus1pt minus4pt

\def\vta{\vartheta}\def\a{\alpha}\def\b{\beta}\def\c{\gamma}
\def\d{\delta}

\def\negenspace{\kern-1.1em}



\def\ltextindent#1{\hbox to \hangindent{#1\hss}\ignorespaces}



\def\sqr#1#2{{\vcenter{\hrule height.#2pt\hbox{\vrule width.#2pt
height#1pt \kern#1pt \vrule width.#2pt}\hrule height.#2pt}}}
\def\square{\mathchoice\sqr64\sqr64\sqr{4.2}3\sqr{3.0}3}

\newcount\refno
\refno=1
\def\y{\the\refno}
\def\myfoot#1{\footnote{$^{(\y)}$}{#1}
                 \advance\refno by 1}


\def\newref{\vskip 1pc 
            \hangindent=2pc
            \hangafter=1
            \noindent}

\def\neq{\hbox{$\,$=\kern-6.5pt /$\,$}}


\def\semidirect{\;{\rlap{$\subset$}\times}\;}



\newcount\secno
\secno=0
\newcount\fmno\def\z{\global\advance\fmno by 1 \the\secno.
                       \the\fmno}
\def\sectio#1{\medbreak\global\advance\secno by 1
                  \fmno=0
     \noindent{\the\secno. {\it #1}}\noindent}
\def\hodge{{}^\star\!}
{\it \hfill Peprint Cologne-thp-1993-H6}
\bigskip\bigskip\bigskip\bigskip
\centerline{\bf TOWARDS COMPLETE INTEGRABILITY OF}
\centerline{\bf TWO DIMENSIONAL POINCAR\'E GAUGE GRAVITY}
\bigskip
\centerline{by}
\bigskip
\centerline{Eckehard W.\ Mielke, Frank Gronwald,
Yuri N.\ Obukhov$^{*\diamond}$,}

\centerline{Romualdo\ Tresguerres$^{**}$, and Friedrich W.\ Hehl}
\bigskip
\bigskip
\centerline{ Institute for Theoretical Physics, University of
Cologne, D--50939 K\"oln}
\centerline{Germany}

\bigskip\bigskip
\bigskip
\centerline{\bf Abstract}

It is shown that gravity on the line can be described by the
two dimensional (2D) Hilbert--Einstein Lagrangian supplemented by a
kinetic term for the coframe and a translational
{\it boundary} term. The resulting model is equivalent to a
Yang--Mills theory of local {\it translations} and frozen Lorentz
gauge degrees.  We will show that this restricted
Poincar\'e gauge model in 2 dimensions is completely integrable.
{\it Exact} wave, charged black hole, and `dilaton' solutions
are then readily found. In vacuum, the integrability of the {\it general}
2D Poincar\'e gauge theory is formally proved along the same line
of reasoning.
\bigskip\bigskip
\bigskip
\bigskip\bigskip

\bigskip\bigskip
\bigskip
\bigskip\bigskip

\vfill


\noindent $^{*})$ Permanent address: Dept.\ of Theor.\ Physics,
Moscow State University, 117234 Moscow, Russia.

\noindent $^{\diamond})$ Alexander von Humboldt Fellow.

\noindent $^{**})$ Permanent address: Consejo Superior de Investigaciones
Cient\'ificas, Serrano 123, Madrid 28006, Spain.
\eject

\sectio{\bf Introduction}

Recently, $2$--dimensional models of gravity
have attracted some attention as a conceptual ``laboratory" for future
studies of gravity
in higher dimensions and
as a basis of string theory. As is well--known, the Hilbert--Einstein
Lagrangian ($s=$ signature)
$$V_{HE}=(-1)^{s}\,{1\over{ 2}}\,
R^{\{\}\alpha\beta}\wedge\eta_{\alpha\beta}\,
\eqno(\z)$$
of GR does not yield any Einstein--type equations in two spacetime
dimensions. (For n=2, no inverse fundamental length $\ell^{-1}$ occurs
in (1.1) as a coupling constant; in $n$ dimensions this factor would
be $\ell^{2-n}$.) Therefore, in the approach of Teitelboim [1] and
Jackiw (TJ model) [2,3], one had to resort to a
dynamical model with constraint in
which the field equation of constant or even vanishing [4]
(scalar) curvature is enforced by means
of a Lagrange multiplier. This {\it teleparallelism constraint}
of the TJ model we will
put in this paper in its proper perspective: Effectively, it yields
a gauge theory of spacetime {\it translations}.

In fact, in n=4 dimensions, a theory of gravity with the constraint of
vanishing Riemann--Cartan curvature $R^{\a\b}$ is known as
teleparallelism
theory [5,6]. It is a gauge theory of local translations [7,8] and
empirically indistinguishable from Einstein's theory of
general relativity. Moreover,
this model remains nontrivial in $n=2$ dimensions and, as it turns out,
has many salient features of the TJ model.In the context of the
string theory, 2D teleparallel models were actually studied previously
[9,10,11].

In this paper we demonstrate the {\it complete integrability} of 2D
teleparallelism in vacuum. In accordance with old mechanical
knowledge on {\it general coordinates} [12], the Lagrange multiplier
$\lambda$ of the constraint $R^{\a\b}=0$ converts into one of the two
coordinates of our exact black hole solution.

The coupling to gauge, scalar, and spinor matter is also studied. It is
a peculiar but common feature of two dimensions that all these fields
have vanishing 2D
spin currents $\tau_{\a\b}$. Thus the material energy--momentum current
is symmetric and covariantly conserved with respect to the Riemannian
connection. This already indicates
that in two dimensions a decoupling from the Lorentz connection
$\Gamma^{\a\b}$ occurs. It considerably facilitates the integrabilty of
gravitationally coupled matter.

Constrained dynamical systems tend to become liberated classically or,
ultimately, by quantum fluctuations. Nevertheless, we will show for
the first time that the {\it general} Poincar\'e gauge (PG) field equations
[13] can be formally
solved in two dimensions. For a {\it complete proof} of integrability, the
gauge field momenta have to be invertible with respect to
torsion $T^\a$  and curvature $R^{\a\b}$. This puts only very mild restrictions
on the
form of the gravitational gauge Lagrangian. As an application, we
demonstrate that the general
$R+ T^2 +R^2$ Lagrangian is completely integrable and has black hole type
solutions [14,15]. In contrast to a previous proof of Katanayev and Volovich
[16], see also Ref.17, we do not have to
rely on specific gauges, such as the conformal gauge for the coframe.

Our paper is organized as follows: In Sect.2 the geometrical structure
of Riemann--Cartan spacetime and some of its peculiarities in two dimensions
is exhibited for both signatures of the metric.
The transition from the Hilbert--Einstein Lagrangian
to teleparallelism is motivated in Sect.3. The resulting field equations
are reduced in Sect.4 in order to facilitate the proof of complete
integrability in Sect.5. In general, we obtain a black hole solution,
whereas a constant torsion leads to the 2D ``gravitational waves"
of Sect.6. The generalization
to charged black holes is straightforward. As shown in Sect. 7, the
gravitationally coupled Yang--Mills system is still completely
integrable. The coupling to scalar fields is notoriously difficult;
nevertheless, an
exact dilaton type solution has been obtained in Sect.8 in the static
massless case. For the Dirac field of Sect.9, a complete decoupling
from the gravitational field equations occurs at least  for
massless fermions. In Sect.10, the conserved Noether currents
are presented such that the identfication of the integration
constant as the  mass of the 2D black hole becomes finally established
transparent in Sect.11. For an arbitrary PG gauge Lagrangian
$V$ the general field equations are formally solved
completely in Sect.12. The former role of the Lagrange multiplier as a
coordinate is now taken over by the momentum conjugate to the curvature.
Lagrangians with invertable gauge field momenta turn out to be
completely integrable. This new result is exemplified for the
$R+ T^2 +R^2$ Lagrangian in Sect.13

\bigskip
\goodbreak

\sectio{\bf Riemann--Cartan spacetime in $n$ and in $2$ dimensions}

The geometrical arena consists of a $n$--dimensional
differentiable manifold $M$ together with a metric
$$g = g_{ij}\,dx^{i}\otimes dx^{j}\, \eqno(\z)$$
and an {\it orthonormal} frame and coframe field, respectively,
$$e_{\alpha} = e^{i}{}_{\alpha}\,\partial_{i}\, ,
\qquad\vartheta^{\beta} = e_{j}{}^{\beta}\,dx^{j}\> .\eqno(\z)$$
They are reciprocal to each other with respect to the
{\it interior product} $\rfloor$, i.e.,
$$e_{\alpha}\rfloor\vartheta^{\beta} =
e^{i}{}_{\alpha}\,e_{i}{}^{\beta} =
\delta_{\alpha}^{\beta}\>.\eqno(\z)$$
In the following, we adhere to the conventions (cf.~Ref.[18]) that
${\alpha}, {\beta}, {\gamma} ... = 0, 1 ...$ $n-1$ are
anholonomic or frame indices, $i, j, k ... = 0, 1 ... n-1$ are
holonomic or world indices, $\partial_{i}$ are the tangent vectors, and
${\wedge}$ denotes the exterior product.

\bigskip
\noindent{\it Table I: Gauge field strengths, matter currents, and
$\eta$--basis}
$$\vbox{\offinterlineskip
\hrule
\halign{&\vrule#&\strut\quad\hfil#\quad\hfil\cr
height2pt&\omit&&\omit&&\omit&&\omit&&\omit&&\omit&&\omit&\cr
& {\bf } && {\bf valuedness} && {\bf p--form } &
& {\bf components } && $n=4$ && $3$ && $2$ &\cr
height2pt&\omit&&\omit&&\omit&&\omit&&\omit&&\omit&&\omit&\cr
\noalign{\hrule}
height2pt&\omit&&\omit&&\omit&&\omit&&\omit&&\omit&&\omit&\cr
& $T^{\alpha}$ && vector && $2$ &
& $n^{2}(n-1)/2$ && $24$ && $9$ && $2$ &\cr
& $R^{\alpha\beta}$ && bivector && $2$ &
& $n^{2}(n-1)^{2}/4$ && $36$ && $9$ && $1$ &\cr
& $\Sigma_{\alpha}$ && vector && $n-1$ &
& $n^2$ && $16$ && $9$ && $4$ &\cr
& $\tau_{\alpha\beta}$ && bivector && $n-1$ &
& $n^{2}(n-1)/2$ && $24$ && $9$ && $2$ &\cr
& $\eta_{\alpha}$ && vector && $n-1$ &
& $n^2$ && $16$ && $9$ && $4$ &\cr
height2pt&\omit&&\omit&&\omit&&\omit&&\omit&&\omit&&\omit&\cr}
\hrule}$$

\bigskip
Anholonomic indices are lowered by means of the metric with signature $s$
with respect to an orthonormal frame:
$$o_{\alpha\beta} =
e^{i}{}_{\alpha}\,e^{j}{}_{\beta}\,g_{ij}\; ,\qquad\qquad
(o_{\alpha\beta}) ={\rm diag}\bigl(\underbrace{-1}_s ,\,\underbrace{1,
\cdots\, ,1}_{n-s}\bigr). \eqno(\z)$$
For s=1 we have Minkowskian and for s=0 Euclidean
signature. In order to be able to relate the pointwise attached tangent
spaces to each other in a differentiable manner, we introduce
a linear {\it connection}
$\Gamma=\Gamma_{\alpha}{}^{\beta}\,L^{\alpha}{}_{\beta}$
with values in the Lie algebra of the $n$--dimensional rotation
group SO(n) or ``Lorentz'' group SO(s, n-s), respectively. In a
holonomic basis, the connection 1--forms can be expanded as
$$\Gamma^{\alpha\beta} =
\Gamma_{i}{}^{\alpha\beta}\,dx^{i}=-\Gamma^{\b\a}.\eqno(\z)$$

Similarly as in the $4$--dimensional Poincar\'e gauge theory
[13], the coframe $\vartheta^{\alpha}$ and the connection
$\Gamma^{\alpha\beta}$ are regarded as {\it gauge potentials}
of local translations and local Lorentz transformations,
respectively. The corresponding field strengths are given by
the {\it torsion} 2--form
$$T^{\alpha} := D\vartheta^{\alpha} = d\vartheta^{\alpha} +
\Gamma_{\beta}{}^{\alpha}\wedge\vartheta^{\beta}={1\over
2}\,T_{ij}{}^\alpha\,dx^i\wedge dx^j\,,\eqno(\z)$$
and the {\it curvature} 2--form
$$R^{\alpha\beta}:= d\Gamma^{\alpha\beta}-
\Gamma^{\alpha\gamma}\wedge\Gamma_{\gamma}{}^{\beta} =
{1\over 2}\,R_{ij}{}^{\a\b}\, dx^{i}\wedge dx^{j}=-R^{\b\a}
\, .\eqno(\z)$$
In an RC--space in an orthonormal frame, the curvature, like the
connection, is antisymmetric in $\a$ and $\b$.
For the irreducible decomposition of torsion and curvature in
exterior form notation, see Ref.~[19].

In order to isolate the Riemannian part of our Riemann--Cartan (RC)
spacetime, we
decompose the Riemann--Cartan connection into the Levi--Civita
connection $\Gamma^{\{\}}_{\alpha\beta}$
and the contortion $1$--form $K_{\alpha\beta}=-K_{\b\a}$:
$$\Gamma_{\alpha\beta} =  \Gamma^{\{\}}_{\alpha\beta} -
K_{\alpha\beta}. \eqno(\z)$$
Algebraically, the contortion is equivalent of the torsion according
to $T^\alpha=\vartheta^
\beta\wedge K_\beta{}^\alpha$.
Then the curvature decomposes into Riemannian and
contortion pieces as follows:
$$R_{\alpha\beta} =  R^{\{\}}_{\alpha\beta} -
DK_{\alpha\beta} +
K_{\alpha}{}^{\gamma}\wedge K_{\gamma\beta}\, .  \eqno(\z)$$

We have developed the general geometrical formalism for arbitrary
n dimensions. However, as we saw already in Table I, for $n=2$,
namely for the {\it 2--dimensional RC-space}, we have 2 translation
and 1 rotation generators. This allows us to introduce a Lie (or
right) duality operation, that is, a duality with respect to the
Lie--algebra indices, which maps a vector into a covector and
vice versa:
$$\psi^{\star}_{\a} :=  \eta_{\a\b}\,\psi^{\b}
\quad\Longleftrightarrow\quad \psi^{\a} =
-(-1)^{s}\, \eta^{\a\b}\,\psi^{\star}_{\a} \, ,  \eqno(\z)$$
The complete antisymmetric tensor is defined by
$\eta_{\a\b} :=\sqrt{|\det o_{\mu\nu}|}\,\epsilon_{\a\b}$, where
$\epsilon_{\a\b}$ is the Levi--Civita symbol normalized to
$\epsilon_{\hat 0\hat 1}=+1$; for details
of the $\eta$--basis, see Appendix A.
For $\psi^\b=\vta^\b$ we get $\vta_\a^{\star}=\eta_\a={}^\ast\vta_\a$.
In the case of a bivector--valued $p$--form
$\psi^{\alpha\beta} =- \psi^{\beta\alpha}$, the Lie dual is defined by
$$\psi^{\star} := {1\over 2} \eta_{\a\b}\,\psi^{\a\b}
\quad\Longleftrightarrow\quad \psi^{\alpha\beta} =
(-1)^{s}\, \eta^{\a\b}\,\psi^{\star} \, ,  \eqno(\z)$$

In two dimensions,  we can appreciably compactify
formulas, according to the notation given in Table II:

\bigskip
\noindent{\it Table II: 2D geometrical objects}
$$\vbox{\offinterlineskip
\hrule
\halign{&\vrule#&\strut\quad\hfil#\quad\hfil\cr
height2pt&\omit&&\omit&&\omit&&\omit&&\omit&&\omit&&\omit&\cr
& {\bf n=2 } && {\bf valuedness} && {\bf p--form } &
& {\bf components} & \cr
height2pt&\omit&&\omit&&\omit&&\omit&\cr
\noalign{\hrule}
height6pt&\omit&&\omit&&\omit&&\omit&&\omit&&\omit&&\omit&\cr
& $\Gamma^{\star} := (1/2)\, \eta_{\a\b} \,\Gamma^
{\a\beta}$ && scalar && $1$ &
& $2$ &\cr
& $t^{\alpha}:=\;^{*}T^{\alpha}$ && vector && $0$ &
& $2$ &\cr
& $T:=e_{\alpha}\rfloor T^{\alpha}$ && scalar && $1$ &
& $2$ &\cr
& $t^2:= o_{\alpha\beta}\, t^{\alpha}\, t^{\beta}$ && scalar && $0$ &
& $1$ &\cr
& $R^{\star} = d\Gamma^{\star} $ &
& scalar && $2$ &
& $1$ &\cr
& $R:=e_{\alpha}\rfloor e_{\beta}\rfloor R^{\alpha\beta}$ &
& scalar && $0$ &
& $1$ &\cr
height6pt&\omit&&\omit&&\omit&&\omit&\cr}
\hrule}$$

For $n=2$
torsion is irreducible and contains only the vector piece
(vector--valued 0--form, see Appendix B for further details)
$$T^{\alpha}:= d\vartheta^{\alpha} +(-1)^{s}\,
\eta^{\alpha}\wedge\Gamma^{\star} =
(-1)^{s}t^{\alpha}\eta\, .\eqno(\z)$$
Since the curvature 2--form has only one irreducible component,
it can be expressed in terms of the curvature scalar:
$$R^{\alpha\beta}=- {1\over 2} R\,\vartheta^{\alpha}\wedge\vartheta^{\beta}
\, .\eqno(\z)$$

Let us confine ourselves to the case $s=1$ up to the end of this section.
The local Lorentz transformations are defined by the $2\times 2$ matrices
$\Lambda_{\beta}{}^{\alpha}(x) \in SO(1,1)$, and read for the basic
gravitational variables
$$
\vartheta^{\prime\alpha}=\Lambda_{\beta}^{-1\alpha}\vartheta^{\beta},\ \ \ \
\Gamma^{\prime \b}_{\a}=\Lambda_{\a}{}^{\c}\Gamma_{\c}{}^{\d}
\Lambda_{\d} ^{-1\b}- \Lambda_{\alpha}{}^{\c}d\Lambda_{\c}^{-1\beta}\,
.\eqno(\z)$$
With respect to the parametrization
$$
\Lambda_{\alpha}{}^{\beta}=\delta^{\beta}_{\alpha}\cosh\omega +
\eta_{ \alpha}{}^{\beta}\sinh\omega\, ,\eqno(\z)
$$
Eqs.(2.14) can be rewritten as follows:
$$
\vartheta^{\prime\alpha}=\vartheta^{\alpha}\cosh\omega -
\eta^{\alpha}\sinh\omega,\eqno(\z)
$$
$$
\Gamma^{\prime\ \beta}_{\alpha}=\Gamma_{ \alpha}{}^{\beta} +
\eta_{\alpha}{}^{\beta}d\omega,\ \ \ \ \ {\rm or}\ \ \ \ \
\Gamma^{\star\prime}=\Gamma^{\star}-d\omega.\eqno(\z)
$$

\bigskip
\goodbreak
\sectio{\bf Teleparallel 2D gravity}

We regard gravity as a Yang--Mills type
gauge theory of translations [7]. In this approach the coframe
$ \vartheta^{\alpha}$ and the torsion
$T^{\alpha} $ are the associated gauge potentials and gauge field
strenghts, respectively. [The intricate details of such a
(generalized) affine gauge approach are spelled out in Ref. [8]. There local
translations are consider as a ``hidden" gauge symmetry such that
no need for a ``central extension" [4] arises.]

In our new model, the two dimensional Hilbert--Einstein Lagrangian
is supplemented by a {\it kinetic term} for the coframe, a
{\it cosmological} term and a {\it boundary} term.
Since  2--forms are constructed solely
from the translational gauge potential $\vartheta^{\alpha}$,
conventional general relativity appears to be rather ``minimally" modified.
Thus we consider, instead of (1.1), the  2D Lagrangian
$$V_{\infty}= V_{HE} + (-1)^{s} {1\over{ 2}}T^{\alpha}\; ^{*} T_{\alpha}
  +\Lambda\eta -(-1)^{s}
d\Bigl( \vartheta^{\alpha}\wedge \;^{*}T_{\alpha}\Bigr)\
- R^{\alpha\beta}\wedge\lambda_{\alpha\beta}\,.
\eqno(\z)$$
The last term depending on the Lagrange multiplier
0--form $\lambda_{\a\b}$ will enforce the constraint
$R^{\alpha\beta}=0$ of vanishing
Riemann--Cartan curvature on the residual Lorentz degrees of freedom.
This corresponds to the {\it teleparallelism condition} and
will replace the Teitelboim--Jackiw constraint of constant or,
recently, vanishing
Riemannian curvature $R^{\{\}\alpha\beta}$.

In order to fully recognize the Yang--Mills type structure of our new
Lagrangian, we employ a geometric identity (see Eq. (5.4) of Ref.[6])
which relates GR to its teleparallelism equivalent GR$_\parallel$ in
$n\geq 2$ dimensions. Since
the torsion 2--form is already irreducible for
$n=2$, this identity reduces rather drastically  to
$$-{1\over{ 2}}R^{\alpha\beta}\;\eta_{\alpha\beta}
+{1\over{ 2}}R^{\{\}\alpha\beta}\;\eta_{\alpha\beta}\equiv
d\Bigl( \vartheta^{\alpha} \;^{*}T_{\alpha}\Bigr)\, .
\eqno(\z)$$

Then, our new Lagrangian (3.1) can be rewritten such that
the total Lagrangian reads
$$L=V_{\infty}+L_{\rm mat}= (-1)^{s} {1\over{ 2}}T^{\alpha}\; ^{*} T_{\alpha}+
\Lambda\,\eta + (-1)^{s} {1\over{ 2}}R^{\alpha\beta}\,\eta_{\alpha\beta}
-R^{\alpha\beta}\,\lambda_{\alpha\beta}+L_{\rm mat}\,.
\eqno(\z)$$
This presentation of the Lagrangian clearly exhibits the leading
Yang--Mills term for the translational field strength, whereas
$\Lambda\eta= (\Lambda/2) \eta_{\alpha\beta}\,\vartheta^{\alpha}
\wedge\vartheta^{\beta}$ {\it formally} corresponds to a mass term
for the coframe. Observe that the Einstein--Cartan term
$(1/2)R^{\alpha\beta}\,\eta_{\alpha\beta}=R^\star=d\Gamma^\star$
is also a boundary term in two dimensions and will, consequently,
 not contribute to the field equations.

\bigskip
\goodbreak
\sectio{\bf Field equations}

The gravitational field equations
resulting from varying (3.3) with respect to $\vartheta^{\alpha}$,
$\Gamma^{\alpha\beta}$ and $\lambda_{\alpha\beta}$ are
$$D\,^{*}T_{\alpha} - {1\over 2}\,(e_{\alpha}\rfloor T^{\beta})\, ^*
T_{\beta} +(-1)^{s} \Lambda\,\eta_{\alpha} =(-1)^{s+1}\Sigma_{\alpha}\; ,\qquad
(1st)
\eqno(\z)$$
$$\,D\lambda_{\a\b} -(-1)^{s}\vartheta_{[\alpha}\,^{*}T_{\beta]}
=\tau_{\alpha\beta}\, ,\qquad (2nd)
\eqno(\z)$$
$$R^{\alpha\beta}=0.\eqno(\z)$$
Observe that the Einstein--Cartan piece in (3.3) does not give a
contribution due to $D\eta_{\alpha\beta}=0$ in two dimensions.
By
relaxing this teleparallelism constraint, one would obtain a more
complicated model (the quadratic theory with Yang--Mills type terms in the
Riemann-Cartan curvature and torsion was analyzed in [16]). We will defer
the analysis of the general theory to Sect. 12.
The right--hand
sides are the current 1--forms $\Sigma_{\alpha}$ and
$\tau_{\alpha\beta}$ of energy--momentum and spin, respectively, of
hypothetical 2--dimensional matter. Eqs. (4.1) and (4.2) represent {\it
four} and {\it two} independent components, respectively.

The integrability condition for the second field equation is
identically satisfied, because
$$DD\lambda_{\alpha\beta}=-
2R_{[\alpha |}{}^{\gamma}\; \lambda_{\gamma |\beta ]}=0
\, \eqno(\z)$$
in a teleparallel (Weitzenb\"ock) spacetime, whereas
$$D(\tau_{\alpha\beta} +(-1)^{s}
\vartheta_{[\alpha}\; \;^{*}T_{\beta ]}) = 0
 \eqno(\z)$$
follows from the `weak' Noether identity (10.2) for matter and
gravitational gauge fields, together with the first
field equation. Thus, the second field equation
determines (non--uniquely) the Lagrange multiplier
$\lambda_{\alpha\beta}$.

In order to simplify the field equations, we substitute
$^{*}T^\a=t^\a$ into the field equations (4.1) and (4.2),
respectively, and recall the formula
$^*(\Phi\wedge\vta_\a)=e_\a\rfloor{}^*\Phi$, which is valid for any
p--form $\Phi$. Moreover note that the torsion square piece in
the Lagrangian is proportional to $t^2$:
$$T^\a{}^*T_\a=(-1)^{s}t^2\,
\eta\,.\eqno(\z)$$
Then we find
$$Dt_\a- (-1)^{s}\left({1\over
2}\,t^2-\Lambda\right)\,\eta_\a=(-1)^{s+1}\Sigma_\a\,,\eqno(\z)$$
$$D\lambda_{\a\b} -(-1)^{s}\vta_{[\a}t_{\b]}=\tau_{\a\b}\,.\eqno(\z)$$
Let us represent the Lagrange multiplier as
$\lambda_{\a\b}=(\lambda/2)\,\eta_{\a\b}$, where $\lambda =
(-1)^{s}2\lambda^{\star}$ according to the notation in (2.10).
Then, in the last equation, it is more economical to switch over
to its Lie dual by multiplying it with $\eta^{\a\b}$:
$${1\over 2}d\lambda +(-1)^{s} {1\over 2}t_\b\eta^\b= \tau^\star\,.\eqno(\z)$$

We do not lose any of its 4 components, if we multiply (4.7)
by $\vta^\b$ from the right and employ the
formula $\eta_\a\wedge \vta^\b =- \delta_\a^\b\,\eta$:
$$D(t_\a\vta^\b )-(-1)^{s}\left(t_\a t^\b-
{1\over 2}\delta_\a^\b\,(t^2 - 2\Lambda)
\right)\,\eta=(-1)^{s+1}\Sigma_\a\wedge\vta^\b\,.\eqno(\z)$$
Thereby, the energy--momentum current of the gravitational field
is nicely represented. The trace of (4.10), on substitution of (B.5) of
Appendix B, reads
$$(-1)^{s}d{}^* T+2\Lambda\,\eta=\Sigma_\a\wedge\vta^\a\,.\eqno(\z)$$
In a similar move, we substitute (B.4) into (4.9):
$$d\lambda + T=2\tau^\star \,.\eqno(\z)$$

A very useful condition for the torsion--squared function $t^2$ can be
derived by transvecting (4.7) with $t^\a$:
$${1\over 2}\,d\,t^2-\left({1\over 2}\,t^2-\Lambda\right)\, T=
(-1)^{s+1}t^\a\,\Sigma_\a\,.\eqno(\z)$$
We eliminate $ T$ by means of (4.12) and find:
$$dt^2+\left(t^2 -2\Lambda\right)\,d\lambda=
2\bigl[(-1)^{s+1}t^\a\Sigma_\a +(t^2- 2\Lambda)\,\tau^\star\bigr]\,.
\eqno(\z)$$

Let us now specialize to the {\it vacuum} field equations.
They read
$$\eqalignno{
dt^2= & -\left(t^2-2\Lambda\right)\,d\lambda\,,&(\z)\cr
d{}^* T= & (-1)^{s+1}2\Lambda\,\eta\,,&(\z)\cr
d\lambda= & -  T\,,&(\z)\cr
R_{\a\b}= & 0\,.&(\z)\cr}$$

In two dimensions, the volume 2-form $\eta$ equips the spacetime
manifold $M$ with a {\it symplectic structure}.
In vacuo, the volume 2-form turns out, via the field equation
$\eta=(-1)^{s+1}d{}^* T /(2\Lambda) =(-1)^{s}d(\vta^\a t_\a)/(2\Lambda)$, to be
an
exact form, as it was conjectured
by  Cangemi and Jackiw (Eq.(2.A7a) of Ref.[4]). Since this volume 2--form
appears explicitly in the Lagrangian (3.1), the
cosmological term in (3.1) turns out to be ``weakly" equivalent to
the boundary term $(-1)^{s+1}d{}^* T /2$. Thus, to some extent,
Machian ideas are realized: In fact, the
total volume $\mu$ of our 2D ``world" is, due to Stokes' theorem, given
by the integral of the dual torsion 1--form along the {\it boundary}:
$$\mu (M) = {(-1)^{s+1}\over{ 2\Lambda}} \int_{\partial M}\;^* T\, .
\eqno(\z)$$
On the other hand the cosmological term  cannot completely
compensate the explicit ``topological" term
 $d(\vta^\a\wedge {}^*T_\a)=-d{}^* T$ in (3.1). Observe also that,
according
to Ref. [4], the 1--form ${}^* T$ seems to be related to a
gauge 1--form $a$ associated with the {\it central extension} of the 2D
Poincar\'e algebra.

In vacuo, $ T$ is also an exact form.
If it were chosen as one basis 1-form, it would be a
natural 1-form, that is, the Lagrange multiplier $\lambda$ could be
interpreted as a coordinate. Such a transmutation of $\lambda$
from a``constraining force" to a generalized coordinate is known from
mechanics [12] and quantum cosmology [20]. However, in our
model, the vacuum field
equations (4.16) and (4.17) impose the Klein--Gordon equation
$$\square\lambda:=(-1)^s\,[{}^*d{}^*d+d{}^*d{}^*]\,\lambda
=(-1)^{s} 2\Lambda\, \eqno(\z)$$
on the ``would--be" coordinate $\lambda$. Fortunately it turns out,
see the next Section, that this is merely a condition on a so far
unspecified  metric function. Thus, Eq. (4.20) resembles the
harmonic gauge condition in 4D general relativity.

Formally, Eq.(4.20) has the solution
$$\lambda=(-1)^{s}2\square^{-1}\;\Lambda, \eqno(\z)$$
such that the constraining part of the Lagrangian (3.1) takes the form
$$(-1)^{s}R{}^{\star}\;\lambda=2R{}^{\star}\square^{-1}\,\Lambda\,. \eqno(\z)$$
By imposing the additional constraint $R=\Lambda$,
one can obtain the ``weak" relation:
$$(-1)^{s}R{}^{\star}\;\lambda\cong 2 R{}^{\star}\,\square^{-1}\,R=
- (R\,\square^{-1}\,R )\eta\,.\eqno(\z)$$
In Riemannian spacetime, this term is easily recognized as
Polyakov's `string inspired' [21] Lagrangian.

\bigskip
\goodbreak
\sectio{\bf  Black hole solution and complete integrability}

The general quadratic Poincar\'e gauge theories in two dimensions
(in absence of matter) are known to be completely integrable [16,17].
Usually this fact is established
with the help of a convenient choice of coordinates, such as the
light-cone or conformal ones. We will demonstrate that the
model under consideration is also
completely integrable. Again the choice of coordinates will be an essential
step, but we will use an approach discussed by Solodukhin [22].

Before integrating the gravitational equations it is worth to notice that
the flat (Minkowski) spacetime arises when both the Riemann-Cartan
curvature and torsion are zero. The former is described by (4.18), but
it is clear that torsion cannot be zero in case of a non--trivial
cosmological
term in (3.1), (4.15), (4.16). Hence the flat Minkowski space-time is not a
vacuum solution of the theory. It is evident also, that in general $t^{2}$
is non-zero --- again the cosmological constant prevents its identical
vanishing.

The vacuum equation (4.15), i.e.
$${dt^2 \over d\lambda} = - t^2 + 2\Lambda,\eqno(\z)$$
can easily be solved for $t^{2}\neq 2\Lambda$
to give the square of torsion as a function of the
Lagrange multiplier
$$ t^{2} =2\Lambda +(-1)^{s} 2M_{0}e^{-\lambda}\,,  \eqno(\z)$$
where $M_0$ denotes an integration constant which, for Minkowskian
signature $s=1$, will later be
identified with the active gravitational mass of the configuration.
Observe that for $M_0=0$ we recover the special solution
$t^{2}=2\Lambda$ which will be analyzed in
the next section.

The field equation (4.17) suggests to interpret the Lagrange multiplier
$\lambda$ as a coordinate, such that
$ T=-d\lambda$ is one leg (`Bein') of an orthogonal coframe.
Let us first construct the frame dual to the coframe. Define the vector field
$$
\xi^\star= - {{t^\a}\over t^2}\,e_{\alpha},\eqno(\z)$$
which is dual to $^{\ast} T$, i.e.,
$$\xi^\star\rfloor\;^{\ast} T = 1,\eqno(\z)$$
cf. (B.9) and (B.10).
In view of (B.12) the equation (4.17) yields
the constancy of the $\lambda$ variable along the vector field
$\xi^\star$,

$$\ell_{\xi^\star}\lambda= \xi^\star\rfloor d\lambda =\xi^{\star}(\lambda)=0\,
,\eqno(\z)$$
where $\ell_{\xi}=\xi\rfloor d + d\rfloor\xi$ is the Lie derivative.
This fact is crucial, since (5.5) allows to introduce a second coordinate,
say $\rho$, defined by the
integral lines of the vector field
$\xi^\star$. In view of (5.5) the ($\lambda, \rho$) system is orthogonal,
and hence the
form $^* T$ should be proportional to $d\rho$, while $ T$ in view
of (4.17) is already proportional to $d\lambda$.

The leg orthogonal to $T$ is ${}^*{ T}$. Thus we introduce the
orthogonal coordinate system $(\lambda , \rho)$. Then
$${}^* T=B(\lambda ,\rho)d\rho\,.\eqno(\z)$$
Because of the orthogonality, there enters no term proportional to
$ d\lambda$. We
substitute the Ansatz (5.6) into (4.16), use the explicit expression (B.20)
of the volume 2--form, and find
$${\partial B(\lambda, \rho)\over\partial\lambda}
=2\Lambda\,{B(\lambda ,\rho)\over t^2}\, .\eqno(\z)$$
The same relation could be obtained from the Klein--Gordon equation
(4.20) for
$\lambda$. Upon integration we obtain:
$$B(\lambda , \rho)=B_0(\rho)\,t^2\,e^{\lambda}\,.\eqno(\z)$$
In terms of the frame (B.15) or (B.19) of Appendix B,
the metric reads explicitly
$$g=o_{\a\b}\,\vartheta^\a\otimes\vartheta^\b=
(-1)^{s}{d\lambda^2\over t^2}+{B^2\,d\rho^2\over t^2}
\,,\eqno(\z)$$
or, after substituting $t^2$ and absorbing $B_0(\rho)$ according to the
coordinate transformation
$d{\widetilde\rho}:=B_0\,d\rho$,
$$g=(-1)^{s}2e^{2\lambda}(M_0e^{-\lambda} +(-1)^{s}\Lambda)\,
d{\widetilde\rho}^2 +
{d\lambda^2 \over 2(M_0e^{-\lambda} +(-1)^{s}\Lambda)}\,.\eqno(\z)$$

It is remarkable to notice that this metric has the
form of the black hole in the two-dimensional dilaton (string motivated)
gravitational theories, widely discussed in the literature (cf. [14,15,23,24
25]). We will see in Sect.11 that the integration constant $M_{0}$ is in fact
related to the mass of this black hole.

Along with the metric (5.10) one can construct the explicit form of
the torsion in coordinates $x^{i}=(x^{0}=\rho,x^{1}=\lambda)$,
$$
T^{i}=e^{i}{}_{\alpha}T^{\alpha}=
{1\over 2}T^{\ \ \ i}_{jk\cdot}dx^{j}\wedge dx^{k}.\eqno(\z)
$$
 From (B.19) one readily obtains the frame
$$
e_{\alpha}=-{t_{\alpha} \over B}\partial_{\rho} +
(-1)^{s}\eta_{\alpha\beta}t^{\beta}\partial_{\lambda},\eqno(\z)
$$
and thus the components of the torsion tensor read
$$
T^{\ \ \ 0}_{01\ \cdot}= (-1)^{s},\ \ \ \
T^{\ \ \ 1}_{01\ \cdot}=0\, .\eqno(\z)
$$

Using the definition of the torsion 2-form (2.6), one can express the
two-dimensional Lorentz connection (2.5), according to Table II,
in the following convenient dual form
$$
\Gamma^{\star}= ({}^\ast d\vartheta^{\alpha})\vartheta_{\alpha}
+ {}^\ast T.\eqno(\z)$$

In order to complete the analysis of the integrability of the model under
consideration one should also study the equation (4.3),(4.18) and verify
that this is fulfilled on the above described solutions. In general this
is a non trivial problem, espesially in presence of matter, see Sect. 7.

Notice that the above solutions completely describe
the behavior of the torsion: equation (5.13) gives the torsion's components
with respect to the local coordinates $(\rho,\lambda)$, while the torsion
square was obtained explicitly in (5.2). Its {\it local Lorentz} (i.e. with
respect to the local orthonormal frame) components seem to remain undetermined,
but this is clearly related to the gauge freedom of the
model, which means that a vector at any point can be arbitrarily rotated
with the help of the local Lorentz transformations (2.14). Let us demonstrate
this
explicitly. Since $t^2$ is the known function of $\lambda$, one can assume
the general ansatz for the local Lorentz torsion components:
$$
t^{\hat 0}=t\sinh{u},\ \ t^{\hat 1}=t\cosh{u},\eqno(\z)
$$
where $t=\sqrt{t^2}$, and $u=u(\rho,\lambda)$ is some function of both, spatial
and time, coordinates which is real for $s=1$ and purely imaginary
 for $s=0$.
Substituting this into (B.19) and differentiating, one
finds
$$
d\vartheta^{\alpha}=(t^{\alpha}{t\over B}[\partial_{\lambda}({B\over t}) +
{1\over t}\partial_{\rho}u] -
\eta^{\alpha}_{\ \beta}t^{\beta}\partial_{\lambda}u)\eta,\eqno(\z)
$$
and hence the dual Lorentz connection (5.14) is calculated to be
$$
\Gamma^{\star}= [B - t\partial_{\lambda}({B\over t})]d\rho - \partial_{\rho}u
d\rho - \partial_{\lambda}ud\lambda=
$$
$$
=[B - \partial_{\lambda}B + {B\over 2}{{\partial_{\lambda}t^2}\over t^2}]d\rho
- du\, .\eqno(\z)
$$
Evidently the last term represents the local Lorentz transformation (2.21) and
can be discarded by choosing in (5.15) the gauge $u=0$.
While calculating the two--dimensional Riemann--Cartan curvature,
one notices
that the last term in (5.17) does not contribute. Therefore (4.18), i.e.,
the vanishing of the curvature 2-form
$$
R^{\star}=d\Gamma^{\star}=0\, ,\eqno(\z)
$$
reduces to the condition
$$
\partial_{\lambda}[B - \partial_{\lambda}B +
{B\over 2}{{\partial_{\lambda}t^2}\over t^2}]=0\eqno(\z)
$$
Using the vacuum solution (5.8), one finds
$$
B - \partial_{\lambda}B + {B\over 2}{{\partial_{\lambda}t^2}\over t^2} =
-{B\over 2}{{\partial_{\lambda}t^2}\over t^2}=(-1)^{s}M_{0}B_{0}.\eqno(\z)
$$
Since Eq. (5.19) holds for the solution (5.10), the proof of
the integrability of the vacuum equations (4.15)--(4.18) is completed.

Let us investigate some of the properties of our
black hole solutions and, in particular, compare these with the dilaton gravity
black holes. In a first step,  one can try to find a new coordinate
$\widetilde\lambda$ such that also $\theta
^{\hat 1}$ can be represented as a natural leg:
$$\theta  ^{\hat 0}=d\widetilde\lambda=d\lambda /
\sqrt{2(M_{0}e^{-\lambda}+(-1)^s\Lambda)}\, .\eqno(\z)$$
Clearly, one has to distingish the different cases: $M_0
\,e^{-\lambda} +(-1)^s\Lambda >0,\;=0\;\;\hbox{or}\;\;<0$.
Here we restrict ourselves to the first case.
Then the coordinate transformation reads:
$$\lambda=-\log\left({(-1)^{s}\Lambda \over M_0}
\sinh^{-2}\bigl(\sqrt{(-1)^s{\Lambda\over 2}}
\widetilde\lambda\bigr)\right)\,.\eqno(\z)$$
Substitution into the metric (5.10) yields
$$g={M_0^2\over 2\Lambda}\,\sinh^2
\bigl(\sqrt{(-1)^s 2\Lambda}\,\widetilde
\lambda\bigr)\,d{\widetilde\rho}^2+ d{\widetilde\lambda}^2\,.\eqno(\z)$$
Then, the further coordinate transformation
$$\widetilde\lambda = {2\over\sqrt{(-1)^s 2\Lambda}} {\rm arctanh}
\left({1\over2}\sqrt{2\Lambda((-1)^s x^2+y^2)}\right).\eqno(\z)$$

$$\widetilde\rho =
\cases{{1\over{ M_0}}{\rm arctan}\left({y\over x}\right) & for $s=0$\cr
        \quad & \ $\quad$ \cr
       {1\over{ M_0}}{\rm arctanh}\left({y\over x}\right) & for
$s=1$\cr}\eqno(\z)$$
converts the metric (5.23) into the explicit {\it conformally flat}
form
$$g= {{(-1)^{s} dx^2 +dy^2}
\over{1-{\Lambda\over 2}\bigl((-1)^s x^2 +y^2\bigr)}}\, .\eqno(\z)$$
\bigskip

\goodbreak\bigskip
\sectio{\bf  Gravitational waves}

In order to exhibit the propagating degrees of freedom of our model, we
consider the vacuum field equations. For {\it nonvanishing}
``cosmological" constant $\Lambda$, we obtain from the 1st field equation
$$D\;^{*}\vartheta^{\alpha} =D\eta^{\alpha} =-
d\log(t^2-2\Lambda )
\wedge\eta^\a \,, \eqno(\z)$$
due to the constraint (4.3), and
$$\square \vartheta^{\alpha} := (-1)^{s}\Bigl[\, ^{*}D\;^{*}D +
\, D\; ^{*}D\,^{*}\Bigr] \vartheta^{\alpha}
=(-1)^{s}\left(t^2 -2\Lambda\right)\vta^\a\, .
\eqno(\z)$$
These gauge--covariant nonlinear {\it Proca type
equations} for the coframe are {\it exact} consequences
of our ``topological" gauge model (3.1) with teleparallelism.

For the special solution $t^2=2\Lambda$, which has been left out in
Sect. 5, these equations simplify to a
wave equation for the coframe:
$$\square\vta^\a=0\, \,.\eqno(\z)$$

We can almost adopt the solution (5.10) of the previous section
for $M_0 =0$, but
use the coordinate freedom to put $B_0= e^{\pm\rho}$.
Then we find the metric
$$g=(-1)^{s}{1\over 2\Lambda}\,d\lambda^2 +
2\Lambda e^{2(\lambda \pm\rho)}\,d\rho^2\, \eqno(\z)$$
of a {\it ``left-- or right moving"} wave solution. It is the
analogue of the {\it plane
fronted} gravitational wave solution
$$g=-d\lambda^2+L^2(\lambda-z)\bigl(e^{2\b(\lambda-z)}dx^2+
e^{-2\b(\lambda-z)}dy^2)+dz^2\eqno(\z)$$
in 4D gravity, cf. [26,p.975].

It can easily be shown that the Cauchy problem for (6.3) is well--posed:
In $2$ dimensions, the coframe
$\vartheta^{\alpha}=e_{j}{}^{\alpha}\, dx^{j}$ has
$2\times 2=4$ components. Two degrees of freedom get fixed by
considering coframes in the conformal gauge
$\vartheta^{\alpha}=\Omega\, dx^{\alpha}$. Moreover, the one
local Lorentz degree of freedom $\Lambda^{\hat 0\hat 1}$
in the transformation formula (2.14) has also to be
subtracted. Then for Minkowskian signature $s=1$, Eq. (6.3) constitutes
a hyperbolic wave equation for the conformal factor $\Omega$  as the
{\it only} remaining dynamical degree of freedom.

Thus our model contains only a massless ``spin--2" mode,
 i.e. a ``topological graviton" in two dimensions. Quantization will be
straightforward. Moreover, by relaxing the teleparallelism constraint,
the extended model with a Yang--Mills type curvature squared term
appears to be renormalizable [27].

\bigskip\goodbreak

\sectio{\bf Charged black holes}

Let us add to our gravitational Lagrangian (3.3) the standard Maxwell
Lagrangian
$$L_{M}= (-1)^{s}\,{1\over 2} F\wedge\;^{*} F,\eqno(\z)$$
where $F=dA$ is the field strength for the abelian gauge potential
$A$. In two dimensions there is no magnetic field: the only component
of the Maxwell tensor describes the electric field along the unique
spatial direction. This is expressed by
introducing the scalar $f$ dual to the Maxwell field strength, i.e.,
$$f :=\;^{*} F ,\qquad  F=(-1)^{s}f\eta\, .\eqno(\z)$$
The energy--momentum current in (4.1) reads
$$ \Sigma_{\alpha}:=e_{\a}\rfloor L_{M} - (-1)^{s}\,
(e_{\alpha}\rfloor F)\wedge\,^* F= -(-1)^{s}\,
{1\over 2}(e_{\alpha}\rfloor F)\wedge\,^* F =-
{1\over 2}f^2 \eta_{\alpha}\,,\eqno(\z)$$
whereas the spin current vanishes, i.e., $\tau_{\alpha\beta}=0\,,$
since the  Lorentz connection does not couple to the Maxwell field.
Observe that the energy--momentum trace, in constrast to four
dimensions, does {\it not} vanish:
$$\vartheta^{\alpha}\wedge\Sigma_{\alpha} =-f^2\eta\,.\eqno(\z)$$
The Maxwell equations are obtained from the variation of (7.1)
with respect
to $A$, and read as usually
$$d\;^* F=df= 0,,\eqno(\z)$$
In two dimensions these are easily integrated to give $f=const=Q$.
Moreover, this constant is indeed the conserved total electric charge.

As a result, the field equations (4.11)--(4.13) are
{\it completely integrable} along the same
lines, and the relevant charged black hole solutions are obtained by the
following simple shift of the cosmological constant:
$$
\Lambda \rightarrow\widetilde\Lambda= \Lambda - {1\over 2}Q^2.\eqno(\z)
$$

It is straightforward to see that this result is also valid for a Yang-Mills
field with an arbitrary gauge group: After replacing $F$ by the
non-abelian Lie algebra--valued form $F^{A}$, and the Lagrangian (7.1) by
$$
L_{YM}= (-1)^{s}\,{1\over 2} F^{A}\wedge\,^* F_{A},
$$
one obtaines, with the aid of (7.6), the same ``charged" black holes
for which
$$Q^2 = f^{A}f_{A},\ \ \ \ f_{A}=^\ast F_{A}.
$$
[Note that $f_{A}$ is not constant in view of the nonlinear
nature of the Yang-Mills equations $D\,^* F_A =d\,f_A +c_{ABC} A^B\,f^C =0$,
but its square is conserved].

\bigskip
\goodbreak
\sectio{\bf Coupling to a scalar field}

\bigskip
Let us now consider a gravitationally coupled scalar field
$\phi$ for which $L_{\rm mat}$ in (3.3) is given by
$$
L_{\phi}= (-1)^{s}\,{1\over 2}d\phi\wedge\ast d\phi+ U\eta\,
.\eqno(\z)
$$
The  potential $U=U(\phi)$ may include the mass term ${1\over 2}m^2\phi^2$ as
well as a nonlinear selfinteraction of scalar matter.
Introducing the notations
$$
\partial_{\alpha}\phi:=e_{\alpha}\rfloor d\phi,\ \ \ (\partial\phi)^2=
g^{\alpha\beta}\partial_{\alpha}\phi\partial_{\beta}\phi,\ \ \
{\cal{L}}:={}^*L_{\phi}={1\over 2}(\partial\phi)^2 + (-1)^{s} U\,,\eqno(\z)
$$
where $\cal{L}$ is just the Lagrangian {\it function},  one finds for the
sources
of the gravitational field
$$
\Sigma_{\alpha}=(-1)^{s+1}[(\partial_{\alpha}\phi)(\partial^{\beta}\phi)
- \delta^{\beta}_{\alpha}{\cal{L}}]\eta_{\beta}\,,\qquad
\tau_{\alpha\beta}=0.\eqno(\z)
$$
The gravitational field equations (4.1)--(4.3) have to be supplemented by the
equation of motion of
the scalar matter:
$$\;^* d\;^* d\phi - {dU \over d\phi} = 0.\eqno(\z)
$$

As a first step towards a solution of the highly nonlinear
system (4.1)--(4.3), (8.4), we will confine ourselves to the
{\it static} case, such that
$$
\partial_{\rho}\phi\sim t^{\alpha}\partial_{\alpha}\phi=0\,.\eqno(\z)
$$
Then, the basic equations (4.15), (4.16) and (4.17)
are modified as follows:
$$
dt^2=\left(-t^2 +2(\Lambda +(-1)^{s} {\cal{L}})\right)d\lambda \,,\eqno(\z)
$$
$$
d\;^* T = (-1)^{s+1} 2(\Lambda + U)\eta ,\eqno(\z)
$$
$$
d\lambda =-  T\,.\eqno(\z)
$$
An important case is obtained, when the scalar field potential
has non-trivial local extrema (normally, minima) ,
say $\phi_{0}\,$, at which
$${dU \over d\phi}=0.\eqno(\z)
$$
Evidently the constant configuration $\phi=const=\phi_{0}$ is a solution of
(8.4) and the remaining gravitational field equations (8.5)--(8.7) are
reduced to those of the ``vacuum case",  except that
the cosmological constant is shifted to
$$
\Lambda\rightarrow\overline{\Lambda}=\Lambda + U(\phi_{0}).\eqno(\z)
$$
The solutions are thus again the black holes (5.10).

For the general case, the scheme of reasoning is the same as in the
Sect.5: the coordinates  $(\rho,\lambda)$ are introduced, such that
(5.6) and (5.9) hold.
However, (4.17) is replaced by (8.6), and this yields for the function
$B(\rho,\lambda)$ the condition
$$
{t^2 \over B}{\partial B \over {\partial \lambda}}
= 2(\Lambda +U).\eqno(\z)
$$
In these coordinates the Klein-Gordon equation for the matter field
reads explicitly
$$
\partial^{2}_{\lambda}\phi +
{1\over B}(\partial_{\lambda}B)\partial_{\lambda}\phi
- {1\over t^2}{dU \over d\phi}=0.\eqno(\z) $$

For $\Lambda= U=0$, similarly as in an exact Einstein--dilaton field
solution in four dimensions [28], the term  ${\cal L}=(1/2)C^2$ effectively
replaces the cosmological constant. In two dimensions this leads to a
a conformally invariant model, for which the exact {\it `dilaton'
solution}
$$\eqalign{\phi &=C\lambda \,, \qquad B_0= B_0(\rho)\,, \cr
  g&={{d\lambda^2}\over{C^2-Ae^{-\lambda}}}+
(-1)^{s}{{B_{0}^2d\rho^2}\over{C^2-Ae^{-\lambda}}} \cr}
 \eqno(\z) $$
can be obtained. In general, the
integration is more difficult mainly due to the necessity to fulfil the
zero curvature constraint (4.18). This imposes on $B$ and $t^2$ the
additional non--trivial constraint (5.18). Consequently, the system (8.5),
(8.10),
(8.11), (5.18) may admit a dilaton black hole type solutions only
for particular forms
of the potential $U(\phi)$. However, it is likely that generic
two-dimensional black holes have no "scalar hair".

\bigskip
\goodbreak
\sectio{\bf {Coupling to Dirac matter}}

In this section and in the Appendix A we consider only Minkowski
spacetime $(s=1)$.
The theory of spinors in two-dimensions can be formally constructed along
the same lines as in $n$ dimensions, see Ref.[29] and
Appendices A and C. However, there are
certain peculiarities due to the abelian nature
of the two-dimensional Lorentz
group. The most unusual feature is the absence of
coupling of the 2D Dirac field to the local Lorentz connection.

Let $L_{\rm mat}$ in (3.3) be now the Dirac Lagrangian
$$
L_{\psi}={i\over 2}(\bar{\psi}\gamma\wedge \;^*  D\psi +
 \;^*   D \bar{\psi}\wedge\gamma\psi)-im\bar{\psi}\psi\eta,\eqno(\z)
$$
where $\gamma:=\gamma_{\alpha}\vartheta^{\alpha}$
is the matrix--valued 1-form of the Dirac algebra in $2D$ satisfying
$\gamma\wedge\gamma=
-2\eta\gamma_{5}$. (For the details on spinors and realization
of the Dirac algebra in two
dimensions see the Appendices A and C.) The covariant
exterior derivative is defined by
$$D \psi=d\psi + \Gamma\psi,\ \ \ \
  D \bar{\psi}=d\bar{\psi}- \Gamma\bar{\psi},\eqno(\z)$$
where
$$\Gamma:= {i\over 4}\Gamma^{\alpha\beta} \sigma_{\alpha\beta} =
{1\over 4}\eta_{\alpha\beta}\Gamma^{\alpha\beta}
\gamma_{5}={1\over 2}\gamma_{5}\Gamma^{\star}\eqno(\z)$$
is the $\overline{SO}(1,1)$--valued connection.

In two dimensions, the connection is not only
abelian but also involves the $\gamma_5\,$--matrix. This implies
that the spin current
$$\tau_{\alpha\beta}:= {{\partial L_{\psi}}\over{\partial
\Gamma^{\alpha\beta}}} ={1\over 8}\overline{\psi}(\;^*\gamma
\sigma_{\alpha\beta} + \sigma_{\alpha\beta}\;^*\gamma )\psi =
-{\imath\over 8}\eta_{\alpha\beta}\overline{\psi}(\;^*\gamma
\gamma_{5} + \gamma_{5}\;^*\gamma )\psi=0
\eqno(\z)$$
vanishes identically on account of (C.4) and (C.5). Consequently,
the Dirac Lagrangian (9.1) reduces to
$$\eqalign{
L_{\psi} &={i\over 2}(\bar{\psi}\gamma\wedge \;^*  d\psi +
 \;^*  d\bar{\psi}\wedge\gamma\psi)+
\Gamma^{\alpha\beta}\wedge\tau_{\alpha\beta}
-im\bar{\psi}\psi\eta \cr
&={i\over 2}(\bar{\psi}\gamma\wedge \;^*  d\psi +
 \;^*  d\bar{\psi}\wedge\gamma\psi)-im\bar{\psi}\psi\eta\,.\cr}
\eqno(\z) $$

Variation of the Dirac action (9.1) with respect to $\bar{\psi}$
yields the Dirac equation
$$
\gamma\wedge \;^*   D \psi +
{1\over 2}\gamma^{\alpha}\eta_{\alpha\beta}T^{\beta}\psi -
m\psi\eta=0\,.\eqno(\z)
$$
Using (9.3) and (5.14) one can verify that (9.6) does not
contain torsion:
the apparent term actually is cancelled by those ``hidden" in the
exterior covariant derivative.
After defining $  D _{\alpha}=e_{\alpha}\rfloor D $, Eq. (9.6) is
equivalent to
$$
\gamma^{\alpha}\,  D _{\alpha}^{\{\}}\psi- m\psi=0.\eqno(\z)
$$
This again proves the absence of any
coupling of 2D Dirac spinors to the local Lorentz connection.

Variation of (9.5) with respect to the coframe yields the
energy-momentum current
$$
\Sigma_{\alpha}\cong -{i\over 2}(\bar{\psi}\gamma^{\beta}
\partial_{\alpha}\psi -
\partial_{\alpha}\bar{\psi}\gamma^{\beta}\psi)\eta_{\beta},\eqno(\z)
$$
where we took into account that the Dirac Lagrangian vanishes ``weakly",
i.e. $L_{\psi} \cong 0$, on account of the
equation of motion (9.6).

Similarly as in the case of the scalar matter, we are not attempting to
find the
general solution, but restrict ourselves to the {\it static} case, when
$$
\partial_{\rho}\psi\sim t^{\alpha}\partial_{\alpha}\psi=0.
\eqno(\z)$$
Then
$$
\Sigma_{\alpha}t^{\alpha}\cong 0,\ \ \ \
\Sigma_{\alpha}\wedge\vartheta^{\alpha}\cong
im\bar{\psi}\psi\eta,\eqno(\z)
$$
where again $L_{\psi}\cong 0$ is used.

Hence the gravitational field equations (4.1)--(4.3) are reduced to the
following system
$$
dt^2 =(2\Lambda-t^2)T,\eqno(\z)
$$
$$
d \;^*  T=(2\Lambda-im\bar{\psi}\psi)\eta,\eqno(\z)
$$
$$
d\lambda=-T,\eqno(\z)
$$
$$
R_{\alpha\beta}=0.\eqno(\z)
$$

In the massless case $m=0$, a  static Dirac field completely
decouples from the gravitational
field equations. Hence, Eqs. (9.11)--(9.14) reduce to the vacuum case
(4.15)--(4.18) and thus give rise to the same black hole and wave
solutions. The massive spinor case will be discussed elsewhere.

\goodbreak
\sectio{\bf Noether identities and conserved currents}

The sources for the gravitational gauge fields are
the  material energy--momentum current $\Sigma_{\alpha}:=
\delta L_{mat}/\delta\vartheta^{\alpha}$ and the spin current
$\tau_{\alpha\beta} := \delta L_{mat}/\delta\Gamma^{\alpha\beta}$,
which are both $(n-1)$--forms in $n$ dimensions. In fact, in 2
dimensions $\Sigma_\alpha$ represents stress -- and this is a
well--known concept of a force distributed over a (1--dimensional
spacelike)
line element. In 4 dimensions, however, $\Sigma_\alpha$ describes
energy--momentum distributed in a (3--dimensional) volume element.
Accordingly, $\Sigma_\alpha$ corresponds to the intuitive notions of
a line--stress and energy--momentum density in 2 and 4 dimensions,
respectively. This convinces us of the correctness of the
interpretation of the $(n-1)$--form $\Sigma_\a$. An analogous
consideration applies to $\tau_{\alpha\beta}$ as spin moment
stress and spin angular momentum density, respectively.

 From local ``Poincar\'e'' invariance $R^n\semidirect SO(s,n-s)$
one finds [5,13], for $n\geq 2$, the {\it 1st} and
the  {\it 2nd Noether identity}
$$D\Sigma_{\alpha}\cong (e_{\alpha}\rfloor
T^{\gamma})\wedge\Sigma_{\gamma} + (e_{\alpha}\rfloor
R^{\gamma\delta})\wedge\tau_{\gamma\delta}
\eqno(\z)$$
and
$$ D\tau_{\alpha\beta}+\vartheta_{[\alpha}\wedge
\Sigma_{\beta]}\cong 0\, .\eqno(\z)$$
These equations having $n$ and $n(n-1)/ 2$ independent
components, respectively, hold only ``weakly'',
denoted by $\cong$, i.e., provided
the matter field equation $\delta L/\delta\psi=0$ is fulfilled.

In two dimensions, the Noether identities can be rewritten as
$$D\Sigma_{\alpha}\cong (-1)^{s}\,\eta_{\alpha}\wedge
(t^{\gamma}\Sigma_{\gamma} -R\tau^{\star})
\eqno(\z)$$
and
$$ d\tau^{\star} - {1\over 2}\eta_{\beta}\wedge
\Sigma^{\beta}\cong 0\, .\eqno(\z)$$
Note that the Lie dual $\tau^{\star}$ of the spin current
is also given by
$$\tau^{\star}={(-1)^s\over 2}{\delta L_{mat}\over \delta
\Gamma^{\star}}\,.\eqno(\z)$$

For spinless matter, the energy--momentum becomes symmetric and covariantly
conserved with respect to the Riemannian connection [30]:
$$\vartheta_{[\alpha}\wedge
\Sigma_{\beta]}\cong 0\,, \qquad\quad  D\Sigma_{\alpha}^{\{\}}\cong 0\, .
\eqno(\z)$$

If a spacetime admits symmetries, we can construct from the Noether
currents a set of {\it invariant} conserved quantities, one for each
symmetry. We consider Killing symmetries, where the vector field $\zeta=
\zeta^\a e_\a$ is a generator of a one--parameter group of
diffeomorphisms. Then it obeys the generalized Killing equation
$$ {\cal L}_{\zeta}\, g = \left(\L_\zeta g_{\alpha\beta}+2g_{\gamma (\alpha}
e_{\beta )}\rfloor\L_\zeta\vartheta ^\gamma \right)\; \vartheta ^\alpha\otimes
\vartheta^\beta =0\;,\qquad\quad
\qquad{\cal L}_{\zeta}\Gamma_{\alpha}{}^{\beta} =0\, , \quad\eqno(\z)$$
where $ {\cal L}_{\zeta}$ is the usual Lie derivative and $\L_\zeta
:=\zeta\rfloor D +
D\zeta\rfloor$ the gauge--covariant version for exterior forms.

As it was shown in Ref. [31], the current 1-form
$$\varepsilon\, {}_{\rm RC}:=\zeta^{\alpha}\,\Sigma_{\alpha} +(e_{\beta}
\rfloor{\buildrel\frown\over{D}}\zeta^{\gamma})\,
\tau^{\beta}{}_{\gamma}\, ,\eqno(\z)$$
involving the exterior covariant derivative ${\buildrel\frown\over{D}}$ with
respect to the transposed connection
$${\buildrel\frown\over{\Gamma}}_{\alpha}{}^{\beta}:=
\Gamma_{\alpha}{}^{\beta} +e_{\alpha}\rfloor T^{\beta}\,,\eqno(\z)$$
is `weakly' such a closed form:
$$ d\, \varepsilon\, _{\rm RC} \cong 0 \, .\eqno(\z)$$
Thus $ \varepsilon\, _{\rm RC}$ is a {\it globally conserved}
energy--momentum
cuurent in the presence of spacetime symmetries

\bigskip
\goodbreak
\sectio{\bf Mass of the black hole solution}

In order to apply these results to our exact black solution, observe that
the metric (5.10) is independent of
the coordinate $\widetilde\rho $. The corresponding
``timelike'' Killing vector field $\partial_{\widetilde\rho}$ can be
immediately expanded in terms of the frame:
$$\xi^{\star}\rfloor \,^* T=1\qquad\Rightarrow\qquad
\zeta=\partial_{\widetilde\rho}=B\xi^\star=-e^{-\lambda}t^\a e_\a\, .
\eqno(\z)$$

The material spin current of our exact solution vanishes [32],
i.e., $\tau_{\alpha\beta}=0$ such that (10.8)
reduces to
$\varepsilon\, {}_{\rm RC}:=\zeta^{\alpha}\,\Sigma_{\alpha}$ . This is,
In fact, a general feature of all the
matter sources considered in this paper: the Yang--Mills bosons, the dilaton
and even a Dirac field. On the other
hand, for a nonzero mass $M_0$,
the material energy--momentum current cannot vanish everywhere, but needs
to have delta type concentration at the origin, i.e.,
$$\Sigma_{\alpha}\sim \delta(0)\,\eta_{\alpha}\, . \eqno(\z)$$

For the derivation of the related weakly conserved current (10.8), it
is convenient to substitute for $\Sigma_{\alpha}$ the field
equation (4.7). Due to (4.9), i.e.,
$t_\a \eta^\a = (-1)^{s} T=(-1)^{s+1}d\lambda$, we easily obtain
 $$\varepsilon\, {}_{\rm RC} =\zeta^{\alpha}\, \Sigma_{\alpha}
   =(-1)^{s}{1\over 2} e^{\lambda}\bigr[dt^2+(t^2 -
2\Lambda)\,\lambda\bigl]\,.\eqno(\z)$$
Since this current can be derived via
$$ \varepsilon\, {}_{\rm RC} =dM\eqno(\z)$$
from the superpotential
$$M=(-1)^s\ e^{\lambda}({1\over 2}t^2-\Lambda)\,,\eqno(\z)$$
the current $\varepsilon\, {}_{\rm RC}$ is conserved, indeed.
If we go to the ``mass shell'', we discover (5.2) with
$$M=M_0\,.\eqno(\z)$$

\bigskip
\goodbreak
\sectio{\bf Integrability of the general PG equations in two dimensions}

For the {\it quadratic} Poincar\'e gauge model (PG) in two
dimensions its complete integrability in vacuum has been
established [16,17,22]. However, these proofs rely on certain choices
of a gauge. In this section, we will extend this result
to the case of the {\it general} $2D$ Poincar\'e gauge theory
{\it without} imposing any gauge condition.

In PG theory the total action of interacting matter and
gravitational gauge fields reads
$$W = \int \Bigl[L( \vartheta^{\alpha},\Psi , D\Psi)+
 V(\vartheta^{\alpha}, T^{\alpha}, R^{\alpha\beta})
\Bigr]
.\eqno(\z)$$
It is a functional of a minimally coupled matter
field $\Psi$, which, in general, may be  a $p$--form,  and of the
geometrical variables $\vartheta^{\alpha}$
and $\Gamma^{\alpha\beta}=-\Gamma^{\beta\alpha}$. Their {\it independent}
variations
yield the following {\it field equations}:
$${{\delta L}\over{\delta\Psi}} ={{\partial L}\over{\partial\Psi}}-
(-1)^{p}\, D{{\partial L}\over{\partial D\Psi}}= 0, \qquad
{\rm (MATTER)}\eqno(\z)$$
$$DH_{\alpha} - E_{\alpha} =
\Sigma_{\alpha}, \qquad {\rm (FIRST)}\eqno(\z)$$
$$DH_{\alpha\beta} - E_{\alpha\beta} =
\tau_{\alpha\beta}. \qquad {\rm (SECOND)}\eqno(\z)$$
Observe that, in two dimensions, the {\it gauge field momenta}
are 0--forms:
$$H_{\alpha} :=-  {{\partial V}\over{\partial
d\vartheta^{\alpha}}} = -  {{\partial V}\over{\partial
T^{\alpha}}},\qquad {\rm and} \qquad H_{\alpha\beta} := -
{{\partial V}\over{\partial d\Gamma^{\alpha\beta}}}= -
{{\partial V}\over{\partial R^{\alpha\beta}}}.
\eqno(\z)$$
The sources of these Yang--Mills type field equations are
the 1--forms of
material {\it  energy--momentum} and {\it spin}, respectively,
  $$\Sigma_{\alpha} := {{\delta L}\over{\delta\vartheta^{\alpha}}}\, ,
\qquad\qquad \tau_{\alpha\beta} :=
{{\delta L}\over{\delta\Gamma^{\alpha\beta}}}\, .\eqno(\z)$$
Due to the universality of the gravitational interaction,
the 1--forms of energy--momentum
$$E_{\alpha} := {{\partial V}\over{\partial
\vartheta^{\alpha}}} = e_{\alpha}\rfloor V +
(e_{\alpha}\rfloor T^{\beta})\wedge H_{\beta} +
(e_{\alpha}\rfloor R^{\beta\gamma})\wedge H_{\beta\gamma} ,  \eqno(\z)$$
and
$$E_{\alpha\beta} :=-\vartheta_{[\alpha}\wedge H_{\beta ]} \eqno(\z)$$
provide a self--coupling of the gravitational gauge field.

The trace of the energy--momentum
current  (12.7), formed with the aid of the coframe $\vartheta^{\alpha}$,
in general gives us back the gauge Lagrangian $V$ amended by Yang--Mills
type terms according to:
$$\vartheta^{\alpha}\wedge E_{\alpha} =2V +
2 T^{\alpha}\wedge H_{\alpha} +
2 R^{\beta\gamma}\wedge H_{\beta\gamma}\, . \eqno(\z)$$

In order to reduce the field equations, we introduce the 1--forms
$$H:=H_{\alpha}\,\vartheta^{\alpha}\>, \qquad\qquad
\;^* H= H_{\alpha}\,\eta^{\alpha}\, . \eqno(\z)$$

In the 1st field equation, we do not lose any of its 4 components,
if we multiply  (12.3)
by $\vta^\b$ from the right:
$$ -D(\vta^\b H_{\alpha})+T^\b H_\a - \vta^\b\wedge E_{\alpha} =
\vta^\b\wedge \Sigma_{\alpha}\,. \eqno(\z)$$

In this presentation the energy current of the gravitational field
comes out nicely. The trace of  (12.11), on substitution of  (12.9) and
 (12.10), reads
$$ -dH -T^\a H_\a - 2V -2 R^{\beta\gamma}\wedge H_{\beta\gamma} =
\vta^\a\wedge \Sigma_{\alpha}\, . \eqno(\z)$$

In the 2nd field equation it is more economical to switch over
to its Lie-dual by multiplying it with $(1/2)\eta^{\a\b}$
and to introduce the notation
$$H^\star ={1\over 2}\eta^{\a \b}H_{\a\b}=:{1\over 2}\kappa\,.\eqno(\z)$$
Then we get
$${1\over 2}d\kappa - {1\over 2}H_\a\eta^\a= \tau^\star\,, \eqno(\z)$$
or, in view of  (12.10),
$$d\kappa -\; ^* H= 2\tau^\star\, . \eqno(\z)$$

Observe that the gravitational energy--momentum current can be
rewritten as
$$E_{\alpha} =(-1)^{s}\,\left[{\cal V} +
t^{\beta}\, H_{\beta} -{1\over 2}R\kappa
\right]\eta_{\alpha}=: (-1)^{s}\,\widetilde{{\cal V}}\eta_{\alpha}\, ,
\eqno(\z)$$
where ${\cal V}:=\;^* V$ is the Lagrangian function (0--form).
Then a very useful condition for the squared translational momentum
$$ H^2:= o^{\alpha\beta}H_{\alpha}\,H_{\beta}  \eqno(\z)$$
can be derived by transvecting the first field equation (12.3)
with $H^\a$:
$${1\over 2}\,d\,H^2- (-1)^{s}\, \widetilde{{\cal V}} \;^* H=
H^\a\,\Sigma_\a\,.\eqno(\z)$$
We eliminate $\;^* H$ by means of  (12.15) and find:
$$dH^2 - (-1)^{s}\,2  \widetilde{{\cal V}}d\kappa  =
2\bigl[H^\a\Sigma_\a- (-1)^{s}\,2 \widetilde{{\cal V}} \,\tau^\star\bigr]\,.
\eqno(\z)$$

Let us now specialize to the {\it vacuum} field equations.
They read
$$\eqalignno{
dH^2=\; & (-1)^{s}\, 2  \widetilde{{\cal V}}d\kappa \,,&(\z)\cr
dH= \;&(-1)^{s}\left[t^{\alpha}H_{\alpha} -
2\widetilde{\cal V}\right ]\eta\,,&(\z)\cr
d\kappa =&\; ^* H\,.&(\z)\cr}$$

Moreover, in our general PG
model,  we can derive from the vacuum field
equations  (12.21) and  (12.20) the Klein--Gordon equation
$$\square\kappa :=(-1)^s\,[{}^*d{}^*d+d{}^*d{}^*]\kappa
=(-1)^{s}\left[2\widetilde{\cal V}-t^{\alpha}H_{\alpha} \right ] \,
\eqno(\z)$$
for the ``would--be" coordinate $\kappa$.

In order to obtain the general solution, one can proceed along the
same lines as in the Sect.5: We introduce
a coordinate system $(\rho,\kappa)$ which is related to the
translational 1--forms (12.10) via
$$H =  Bd\rho\, ,\qquad\qquad \;^* H=d\kappa,\eqno(\z)$$
with some function $B(\rho,\kappa)$. Similarly as in the
teleparallel case, the volume 2--form is, for $H^2\neq 0$,  given by
$$\eta=-{B\over{H^2}}\, d\kappa\wedge d\rho\, ,\eqno(\z)$$
cf. (B.20) with the torsion 1--form being replaced by
$\;^* T\rightarrow (-1)^{s} H$.

Insertion of this ansatz  into (12.21) together with (12.20) yields
$${\partial  \over \partial\kappa}\ln B =
(-1)^{s}\left[2\widetilde{\cal V}-t^{\alpha}H_{\alpha} \right ]
{1\over{H^2}}= {\partial  \over \partial\kappa}\ln H^2 -
(-)^{s} {{t^\a H_\a}\over{ H^{2}}}\,.\eqno(\z)$$
A formal integration of (12.26) straightforwardly leads to the
solution
$$B=B_{0}(\rho)H^{2}\exp \left((-1)^{s+1}\int d\kappa
{{t^{\alpha}H_{\alpha}}\over{H^2}}
\right)\,,\eqno(\z)$$
where again $B_{0}(\rho)$ is an arbitrary function only of $\rho$.

Let us conclude with several remarks on the integration of the vacuum
field equations (12.3), (12.4) and (12.20)--(12.22). First of all, we
will treat the case $H^2 =0$.

In two dimensions curvature has only one non-trivial component, namely
the curvature scalar $R$. Thus, in view of (2.17), the general
gravitational action (12.1) has the form
$$
V(\vartheta^\alpha , T^\alpha , R^{\alpha\beta})=
V(\vartheta^\alpha , T^\alpha , R)\,.\eqno(\z)
$$
The gravitational gauge field momentum (12.13), on account of (12.5),
can then be rewritten as
$$
\kappa = 2\,{\partial {\cal V} \over \partial R}\,.\eqno(\z)
$$
For $H^2 =0$, Eq.(12.20) yields $\tilde{\cal
V}{=0}$, and, by (12.16), $E_\alpha =0$. Hence in vacuum the 1st field equation
(12.3) degenerates to $DH_{\alpha}=0$. The integrability
condition of this equation is the vanishing of the curvature:
$R_{\alpha\beta}=0$. If this is fulfilled, we actually return to the
teleparallel case, which was analyzed in detail in previous sections. However,
in general, $R_{\alpha\beta}\not=0$. Thus $H_{\alpha}=0$. Then, in
combination with the equation $\tilde{\cal V}{=0}$,we find
${\cal V} = R\kappa/2 = R\,(\partial{\cal V}/\partial
R)$. Summarizing, one is left with two {\it algebraic} equations for
curvature and torsion $$ H_{\alpha}=0,\ \ \ \ \ {\cal
V}-R{\partial{\cal V}\over \partial R}=0,\eqno(\z) $$ the roots of
which yield {\it constant} values for $R$ and $T$.

Let us now turn to the general case with $H^2\not=0$.
Since $R$ is scalar, it is clear from (12.28) that, modulo boundary
terms, torsion can only appear in $V$ in the form of the scalar $t^2$, i.e.
$$
V(\vartheta^\alpha , T^\alpha , R)=
V(\vartheta^\alpha , t^2 , R)=(-1)^{s}\,{\cal V}{(t^2 , R)\eta}\,.\eqno(\z)
$$
Hence the relevant translational momentum reads
$$
H_{\alpha}=-2\,{\partial{\cal V}\over \partial t^2}\,t_\alpha
:=P(t^2,R)t_{\alpha}\,.\eqno(\z)
$$
Together with $\kappa=\kappa(t^2,R)$, this function
plays a decisive role in the {\it formal} integration of the system
(12.20)--(12.22). Indeed, since $t^{\alpha}H_{\alpha}=Pt^2$ and $H^2 =P^2 t^2$,
one recognizes (12.20) as a well-posed equation which involves one
dependent (say $t^2$) and one independent (say $R$) variable. Provided $V$,
and hence $P$,
is smooth, the solution of this first order ordinary differential equation
always exists, thus completing our formal demonstration of the integrability
of the general two-dimensional Poincar\'e gauge theory. Remarkably, the
complete vacuum solution (if $H^2\not=0$) is again of the black hole type
with the metric
$$
g=(-1)^{s}{d\kappa^2 \over H^2} +
H^2 \exp\Big( (-1)^{s+1}\int {t^{\alpha}H_{\alpha} \over H^2} \Big)
d\rho^2,\eqno(\z)
$$
even if it is
more complicated than (5.10). Torsion and curvature for our solution
are obtained by inverting the definitions in (12.5) of the gauge field
momenta, or equivalently, by inverting the relations $\kappa=\kappa(t^2,R),
P=P(t^2,R)\rightarrow t^2=t^2(\kappa,P), R=R(\kappa,P)$. For the solution
to be unique, one must assume the relevant
Hessian (${\partial^2{\cal V}\over \partial R\partial R}, {\partial^2{\cal V}
\over \partial t^2 \partial t^2}$) to be non-degenerate. It is
straightforward to derive from (5.14) the curvature scalar of
our general solution:
$$
R=-(-1)^{s}{H^2 \over B}{\partial \over \partial\kappa}\Big({B\over H^2}
{\partial\over \partial\kappa}H^2 \Big).\eqno(\z)
$$
\bigskip\goodbreak
\sectio{\bf Complete integrability of quadratic PG Lagrangians in
two dimensions}

Let us apply these results of the general Lagrangian to a specific
example, namely to a Lagrangian with terms quadratic in
torsion and curvature. Since the torsion and curvature posses only one
irreducible piece, respectively, the most general quadratic
(parity conserving) Lagrangian reads:
$$L=\, (-1)^s \left( \, {a\over 2}\, T_\alpha {}^*T^\alpha
+{1\over 2}\, R^{\alpha\beta}\eta _{\alpha\beta}+{b\over 2}\,
R_{\alpha\beta }{}^*R^{\alpha\beta }\right)+\Lambda\,\eta
+L_{\hbox{mat}}\,.\eqno(\z)$$
Following the prescriptions (12.5), (12.7), and (12.8), respectively,
we calculate from (13.1) the  gauge field momenta
$$H_\alpha =-(-1)^s\, a\, t_\alpha \, \eqno(\z)$$
and
$$H_{\alpha\beta } =-{1\over2}(-1)^s\left( 1-bR\right)\eta _{\alpha\beta }
\,,\eqno(\z)$$
as well as the gravitational energy--momentum current
$$E_\alpha =-\left( {a\over2}t^2+(-1)^s{b\over4}\, R^2-\Lambda
\right)\eta _\alpha \,,\eqno(\z)$$
and the gravitational spin current
$$E_{\alpha\beta }=\, (-1)^s a\,\vartheta _{[\alpha }\, t_{\beta ]}\,
\,.\eqno(\z)$$
Then the vacuum field equations (12.20)--(12.22) in condensed form read:
$$a^2 d\, t^2 =-\left( at^2+(-1)^s{b\over 2}\, R^2 -2\Lambda
\right) d\,\kappa\,,\eqno(\z)$$
$$a\, d\, {}^*T =\, \left({b\over 2}\,
R^2-(-1)^{s}2\Lambda\right)\eta\,,\eqno(\z)$$
and
$$d\,\kappa =\, b\, d\, R=-a\, T \,.\eqno(\z)$$

In the special case $t^2=0$, we are led  a space
of constant Riemannian curvature as in the TJ model:
$$T^\a=0\, ,\qquad \quad R^2= (-1)^{s} {4\over b}\Lambda\, .\eqno(\z)$$
Otherwise, we find from (13.6) and (13.8) by integration
$$t^2 =(-1)^s\left(\, 2M_0 e^{-(bR/a) }- {b\over{2a}}R^2 +R
+(-1)^{s}{{2\Lambda }\over a}-{a\over b}\right) \,.\eqno(\z)$$
According to (13.8), the torsion 1--form $T $ is again an exact form:
$$T =d\left(-{b\over a}\, R\right) \,.\eqno(\z)$$
Thus we can repeat the reasoning of Sect.5 and regard $dR$
as one natural leg. Then $R$ is the associate ``coordinate" such that
$^*T $ is orthogonal to $T $ , implying again the ansatz:
$$^*T =:\, B(\rho ,R)\, d\,\rho \,.\eqno(\z)$$

Following the steps done in (12.25)--(12.26) with $^*H=-a\, T $
and $H=\,(-1)^s a\,{}^*T $, the unknown function $B$ in (13.12) turns out
to be
$$B(\rho ,R)=\, B_0(\rho)t^2 e^{b R/a}\,.\eqno(\z)$$
For the black hole solution, we can put
$B_0=\, 1 $ without loss of generality and, eventually,
obtain the following new 1-form basis (cf. (B.15) of Appendix B):
$$\eqalign{\theta ^{\hat 0} :=&\,{{T}\over{\sqrt{t^2 }}}=\,
- {b\over{a\sqrt{t^2}}}\, d\, R\, , \cr
\theta ^{\hat 1} :=&\,{{^*T}\over{\sqrt{t^2 }}}=\,\sqrt{t^2}
e^{bR/a}\, d\,\rho\,, \cr }\eqno(\z)$$
with the square of the torsion components given by (13.12).
Accordingly, the torsion components fulfil the relation
$$T^{\hat 0}=0\, ,\qquad\quad T^{\hat 1}=-\sqrt{t^{2}}\,
\theta^{\hat 0}\wedge \theta^{\hat 1}\,.\eqno(\z)$$

Since the metric is again given by (12.28), the proof of integrability
of the  general quadratic 2D PG
model is {\it formally} completed.

This was first done in Ref.[22],
but the following essential point was not explicitly demonstrated:
Compared to teleparallelism model,
where the Lagrange multiplier is an {\it independent} field which can
``transmute" freely to a coordinate, there seems to be a
catch in the case of the general theory. The scalar curvature $R$, regarded as
a
``coordinate", may still keep a remembrance of its origin as a
derivative of the Riemann--Cartan connection. For our exact
solution, the connection 1--form $\Gamma^\star$  contains
the leg $dR$ in its expansion. Thus we have to face a
highly {\it implicit} interrelation between the curvature
$R^\star =d\Gamma^{\star}$ and the {\it formal} coordinate $R$.
Fortunately, one can show the selfconsistency of our scheme: inserting
(13.9) and (13.12) into (12.34), one can verify that
the scalar curvature $R(\rho, R)$ is indeed equal to
$R$ regarded as the coordinate. [We checked this also
also with the aid of the EXCALC package of REDUCE [33]).
This finally concludes the proof of {\it complete integrabilty} of
the $R +T^2 +R^2$ model.

\bigskip
\goodbreak
{\bf Appendix A: (Anti--)selfdual basis for exterior forms
in 2 dimensions}

The symbol $\wedge$ denotes the exterior product of forms, the symbol
$\rfloor$ the interior product of a vector with a
form and $\hodge$ the Hodge star (or left dual) operator, which
maps a p--form into a $(2-p)$--form. It has the property that
$$\hodge\,\hodge\Phi^{(p)}=(-1)^{p(2-p)+s}\Phi^{(p)},\eqno(A.1)$$
where p is the degree of the form $\Phi$.

The volume $2$--form is defined by
$$\eta := {{1}\over{2}}\,\eta_{\alpha\b}\,
\vartheta^{\a}\wedge\vartheta^{\b},
\eqno(A.2)$$
where $\eta_{\a\b}:=\sqrt{|\det o_{\mu\nu}|}\,
\epsilon_{\a\b}$, and
$\epsilon_{\a\b}$ is the Levi--Civita symbol normalized to $\epsilon_{\hat 0
\hat 1}=+1$.
Together with $\eta$, the following forms span a basis for the
algebra of arbitrary p--forms in n dimensions,
$$\eqalign{\eta_{\a}&:=
e_{\a}\rfloor\eta =\, ^{*}\vartheta_{\a}\,,\cr
\eta_{\a\b}&:= e_{\b}\rfloor \eta_{\a} =
\,^{*}(\vartheta_{\a}
\wedge\vartheta_{\b})\,.\cr} \eqno(A.3)$$
We will call the forms
$$\{\eta,\,\eta_{\a},\,\eta_{\a\b}\}\eqno(A.4)$$
the $\eta$--basis of the 2--dimensional space.
In 2 dimensions $\eta_{\a\b}$ is a $0$--form which we took
for the definition
of the Lie dual in (2.10). For the inversion of the Lie dual, we
have to use the 2--dimensional relation
$$\eta_{\a\c}\eta^{\b\c}= (-1)^s\,\,\delta_\a^\b\,.\eqno(A.5)$$

 From Table I we recognize that 2--forms are of central importance
in 2--dimensional gravity. This is also true for
$\eta_\a$, which is Lie dual to the 1--form $\vta^\b$. Indeed,
$$\eta_\a=\eta_{\a\b}\,\vta^\b=\,^*\vta_\a\,.\eqno(A.6)$$

The exterior product of the coframe with the $\eta$--basis
satisfies the following relations:
$$\eqalign{
  \vartheta^\c\wedge\eta_{\a}&=\delta^\c_{\a}\,\eta\,,\cr
  \vartheta^\c\,\eta_{\a\b}&=-\delta^\c_{\a}\,
   \eta_{\b}+\delta^\c_{\b}\,\eta_{\a}\,,\cr}\eqno(A.7)$$
which imply, in particular, that
$$\eta ={1\over 2}\,  \vartheta^\b\wedge\eta_{\b}\,.\eqno(A.8)$$
Differentiating the $\eta$'s yields:
$$\eqalign{D\eta_{\a}&=T^\c\wedge\eta_{\a\c}\,,\cr
           D\eta_{\a\b}&=0\,.\cr}\eqno(A.9)$$

For $s=1$, the Lorentz transformation (2.15) suggests to introduce
the 1--forms which are irreducible with respect to the
connected component of the Lorentz group:
$${\buildrel (\pm) \over \sigma} {}^{\alpha}=
\vartheta^{\alpha}\pm\eta^{\alpha}.\eqno(A.10)$$
These forms are self-- and anti--self dual
$$\;^* {\buildrel (\pm) \over \sigma} {}^{\alpha}=
\pm {\buildrel (\pm) \over \sigma} {}^{\alpha}\, ,\eqno(A.11)
$$
and satisfy the relations
$$
{\buildrel (\pm) \over \sigma} {}^{\alpha}\wedge
{\buildrel (\pm) \over \sigma} {}^{\beta}=0,\ \ \ \ \
{\buildrel (\pm) \over \sigma} {}^{\alpha}\wedge
{\buildrel (\mp) \over \sigma} {}^{\beta}=
2\eta(\mp o^{\alpha\beta} -\eta^{\alpha\beta}).\eqno(A.12)
$$
Under the $SO_{\circ}(1,1)$ transformations (2.20), these objects
simply transform
as
$$
{\buildrel (\pm) \over \sigma} {}^{\prime\alpha}=
e^{\pm\omega}\,{\buildrel (\pm) \over \sigma} {}^{\alpha}\,.\eqno(A.13)
$$
This became manifest in the theory of spinors in two dimensions (see
Sect.9). Eqs. (A.11) show that each $\sigma$--form  actually has
only one independent component which can be denoted as
$$
{\buildrel (\pm) \over \sigma}={\buildrel (\pm) \over \sigma} {^{\hat{0}}} .
\eqno(A.14)
$$
For $2D$ spinors, these are the "generalized Pauli matrices".
\bigskip
\goodbreak

{\bf Appendix B: The many faces of 2-dimensional torsion}

We recognize that the translational field momentum (12.5), i.e.
$H^\a=-\partial V/\partial T^\a=(-1)^{s+1}{}^*T^\a$,
is, in 2 dimensions, a 0-form. In 2 dimensions, the torsion has 2
independent components. And so has its Hodge dual, i.e.,
$$t^\a:={}^*T^\a\qquad\hbox{with}\qquad T^\a=(-1)^s\,{}^*t^\a\,,\eqno(B.1)$$
according to (A.1). Thus, instead of $T^\a$, we can express the field
equations in terms of the equivalent
$t^\a=H^\a$. This is more convenient, since a
0-form can be handled more easily than a 2-form.
The 2-form $T^\a$ can also be
developed with respect to the volume 2-form $\eta$:
$$T^\a= (-1)^{s}{}^*t^\a=(-1)^{s}{}^*t^\a\,1=(-1)^{s}{}t^\a{}^*1=
(-1)^{s}t^\a\,\eta\,.\eqno(B.2)$$

The torsion 2-form $T^\a$ is not only fully contained in the 0-form
$t^\a$, but also in a 1-form $T$. This comes about as follows: In
n-dimensions, the torsion can be decomposed into 3 irreducible pieces,
a tensor, a vector, and an axial vector piece, see [19]. In 2
dimensions torsion is irreducible and only the (co-)vector piece
survives:
$${}^{(2)}T^\a:=\vta^\a\wedge(e_\b\rfloor T^\b)=\vta^\a\wedge T
\qquad\hbox{with}\qquad T:=e_\b\rfloor T^\b\,.\eqno(B.3)$$
The 1-form $T$ can be expressed in terms of $t_\a$ as follows:
$$T =e_\b\rfloor T^\b=(-1)^{s}t^\b\,e_\b\rfloor\eta=
(-1)^{s}t_\b\eta^\b\,.\eqno(B.4)$$
Due to $e_\a \rfloor \,^* \Phi =\,^*(\Phi\wedge \vartheta_{\alpha})$,
its dual reads
$${}^* T={}^*(e_\b\rfloor T^\b)=(-1)^{s}{}^*(e_\b\rfloor{}^*t^\b)=
(-1)^{s}\;{}^{**}(t^\b\vta_\b)=-t_\b\,\vta^\b\,.\eqno(B.5)$$
Now it is easy to show that the vector piece of the torsion coincides
with the total torsion:
$${}^{(2)}T^\a=\vta^\a\wedge T=(-1)^{s}\vta^\a\wedge t_\b\eta^\b=
(-1)^{s}t_\b\vta^\a\wedge\eta^\b=(-1)^{s}t^\a\eta=T^\a\,.\eqno(B.6)$$

Accordingly, we recognize that the torsion can alternatively presented
by the 0-form $t^\a$, the two 1-forms $T$ or $^* T$, or by the
standard 2-form $T^\a$:
$$t^\a:={}^*T^\a\,,\qquad T:=e_\a\rfloor T^\a\,,\qquad
{}^* T= -\vta^\b{}^*T_\b\,,\qquad T^\a\,, \eqno(B.7)$$
with
$$T^\a=(-1)^{s}{}^*t^\a=(-1)^{s}t^\a\eta=\vta^\a\wedge T=(-1)^{s}
\eta^\a\wedge{}^* T\,.
 \eqno(B.8)$$
The set $\{t^\a, T,{}^* T,T^\a\}$ of equivalent torsion forms
turned out to be very useful.

For the presentation of exact 2D solutions it is rather convenient to
introduce, instead of $\vta^\a$ and $e_\b$, quite generally the new
coframe $\{ T,{}^* T\}$ together with its dual vectors
$\{\xi,\xi^\star\}$. By duality we have:
$$ \eqalignno{\xi\rfloor T=1\qquad\Rightarrow\qquad &
(-1)^{s}\xi^\a e_\a\rfloor
(t_\b\eta^\b )=(-1)^{s}\xi_\a\, t_\b\,\eta^{\b\a}=1\,, &  (B.9)    \cr
\xi^\star\rfloor{}^* T=1\qquad\Rightarrow\qquad &-\xi^{\star\a}e_\a\rfloor
(t_\b\vta^\b )=-\xi^{\star\a}\,t_\a=1\,. &  (B.10)  \cr}$$
Apart from singular points, the condition $t^2\neq 0$ holds as a result of
the field equation.
Then we find
$$\xi=-{\eta^{\a\b}\,t_\b\over t^2}\,e_\a\,,\qquad
\xi^\star=-{t^\a\over t^2}\,e_\a\,. \eqno(B.11)$$
Furthermore we can check
$$\xi\rfloor{}^* T=0\qquad\hbox{and}\qquad \xi^\star\rfloor T
=0\,. \eqno(B.12)$$

As a final proof that $ T$ and ${}^* T$ formally span a coframe,
we display the orthogonality of $\xi$ and $\xi^\star$ with the help
of the metric $g$. Since the $e_\a$ are orthonormal, we have
$$g(\xi,\xi^\star)=-{\eta^{\a\b}t_\b t^\c\over t^4}\,g(e_\a,e_\c)
=-{\eta^{\a\b}t_\a t_\b\over t^4}=0\,. \eqno(B.13)$$
The vectors $\{\xi,\xi^\star\}$ are not of unit length, rather
$$\xi^2:=g(\xi,\xi)=(-1)^{s}\xi^{\star 2}:=
(-1)^{s}g(\xi^{\star},\xi^{\star})={{(-1)^{s}}\over t^2}
\,. \eqno(B.14)$$
Consequently, the new coframe
$$\theta^\a=\{\theta^{\hat{0}},\theta^{\hat{1}}\}:=
\Bigl\{{ T\over\sqrt{t^2}},{{}^* T\over\sqrt{t^2}}\Bigr\}=
\Bigl\{(-1)^{s}{t_\b\over\sqrt{t^2}}\,{}^*\vta^\b,-{t_\b\over\sqrt{t^2}}\,
\vta^\b\Bigr\}\,, \eqno(B.15)$$
is {\it orthonormal}, but not the orthogonal system
$\{ T,{}^* T\}$. The dual frame reads
$$E_\a=\{E_{\hat{0}},E_{\hat{1}}\}:= \Bigl\{{\eta^{\b\c}\,t_\c
\over\sqrt{t^2}}\,e_\b\,,-{t^\b\over\sqrt{t^2}}\,e_\b\Bigr\}\,,
\eqno(B.16)$$
that is,
$$E_\a\rfloor\theta^\b=\delta_\a^\b\qquad
\hbox{and}\qquad g(E_\a,E_\b)=o_{\a\b} \eqno(B.17)$$
with
$$E_\a=E^i{}_\a\,{\partial\over\partial X^i}\qquad\hbox{and}
\qquad\theta^\b=E_i{}^\b\, dX^i\,, \eqno(B.18)$$
where $X^i$ are some (holonomic) coordinates.

A 2-dimensional Lorentz transformation depends only on one parameter.
 From (B.16) we can read off its inverse
$$\vta^\a=-{\eta^{\a\b}\,t_\b\over\sqrt{t^2}}\,\Bigl({ T\over\sqrt{t^2}}
\Bigr)-{t^\a\over\sqrt{t^2}}\,\Bigl({{}^* T \over\sqrt{t^2} }
\Bigr)\,. \eqno(B.19)$$
The volume 2-form can also be expressed in terms of the
coframe (B.15):
$$\eta:=\theta^{\hat{0}}\wedge\theta^{\hat{1}}={{(-1)^{s}}\over
t^2}\, T\wedge{}^* T\,. \eqno(B.20)$$

\bigskip\goodbreak

\sectio{\bf Appendix C: Spinors in two dimensions}

Dirac spinors in two dimensions have two (complex) components,
$$
\psi= \pmatrix{ \psi_1 \cr \psi_2 },\eqno(C.1)
$$
and, as usually, the spinor space at any point of the space-time manifold
is related to the tangent space at this point via the spin-tensor objects:
the Dirac and the Pauli matrices.

The Dirac matrices $\gamma^\alpha$ satisfy the standard relations
$$
\gamma^{\alpha}\gamma^{\beta}+\gamma^{\beta}\gamma^{\alpha}=2o^{\alpha\beta},
\eqno(C.2)
$$
and in $2D$ these $2\times 2$ matrices can be chosen to be real.

Further elements of the 2D Clifford algebra are the $\gamma_5$ matrix and
the $\overline{SO}(1,1)$--generator $\sigma_{\alpha\beta}$ which are
defined by
$$\gamma_5 := {1\over 2}\eta_{\alpha\beta}\gamma^{\alpha}\gamma^{\beta}\, ,
\qquad\quad \sigma_{\alpha\beta} :={\imath\over 2}\,
[\gamma_{\alpha}\, ,\gamma_{\beta}]\,.
\eqno(C.3) $$

 From (C.2) and (C.3) one can derive the useful relations:
$$
\gamma_{\alpha}\gamma^{\beta}\gamma^{\alpha}=0,\ \ \ \ \
\gamma_5 \gamma^{\alpha} + \gamma^{\alpha}\gamma_5 =0,\ \ \ \ \
(\gamma_5)^2 = 1\,,\eqno(C.4)
$$
and
$$
\gamma^{\alpha}\gamma_5 = \eta^{\alpha\beta}\gamma_{\beta}\, ,\qquad
[\gamma_{\alpha}\, ,\gamma_{\beta}]=-2\eta_{\alpha\beta}\gamma_{5}\,.
\eqno(C.5)$$

If we introduce the matrix-valued 1--form
$$
\gamma=\gamma_{\alpha}\vartheta^{\alpha}\, ,\eqno(C.6)
$$
Eqs. (C.2)--(C.5) now can be rewritten in Clifford algebra--valued
exterior forms as
$$
\gamma\otimes\gamma=g,\ \ \ \ \
\gamma\wedge\gamma=-2\gamma_5 \eta,\ \ \ \ \
\;^*\gamma =\gamma_5 \gamma.\eqno(C.7)
$$

The action of the gauge (local Lorentz) group on spinors is given by
$$
\psi\rightarrow\psi^\prime = S\psi,\ \ \ \ \
\bar{\psi}\rightarrow\bar{\psi}^{\prime}=\bar{\psi}S^{-1},\eqno(C.8)
$$
where the Dirac conjugation is defined as
$$
\bar{\psi}:=\psi^{\dagger}\gamma^{\hat{0}}.\eqno(C.9)
$$

Eqs. (C.2) and (C.7) relate the metric structure on a spacetime to the
spinor space. Then the Lorentz transformation are converted via the covering
homomorphism $\overline{SO}(1,1)\approx SO(1,1)$ to the similarity
transformations
$$
\gamma^\prime =S\gamma S^{-1},\eqno(C.10)
$$
of the $\gamma$--matrices, where
$\gamma^\prime=\gamma_{\alpha}\vartheta^{\prime\alpha}$. Substituting
(2.14) and using the explicit form of the local Lorentz rotations (2.15),
one finds
$$
S=\exp({\omega \over 2}\gamma_5)=
\cosh({\omega \over 2}) + \gamma_5 \sinh({\omega \over 2}).\eqno(C.11)
$$

This completes the definition of a {\it spinor algebra} on a $2D$ manifold.
The next step is to develop the spinor {\it analysis}, and the
central point is the notion of the so-called
spinor covariant derivative $D$.
The formal definition of the spinor covariant derivative is given by (9.2),
where the connection 1-form, due to the covering homomorphism, has
the usual transformation law
$$\Gamma\rightarrow \Gamma{^\prime}=
S\Gamma S^{-1} + SdS^{-1}\; .\eqno(C.12)
$$
The explicit form of the  connection (9.3) is obtained from the
natural assumption that spinor bilinear terms behave covariantly, that is
$\bar{\psi}\psi,\ \ \bar{\psi}\gamma\psi \ \ {\rm and} \ \
\bar{\psi}\gamma\wedge\gamma\psi$ are the $0-,1-$ and
$2$--forms, respectively,
on the space-time manifold. This is equivalent to the condition
$$
{\cal D} \gamma^{\alpha} = d\gamma^\alpha +
\Gamma^{\ \alpha}_{\beta}\gamma^\beta +
[ \Gamma,\gamma^\alpha ]=0\, ,\eqno(C.13)
$$
for which the explicit solution is just (9.3).

The concrete realization of the Clifford algebra (C.2) and(C.7)
on a 2D manifold
is easily achieved in terms of the $1\times 1$ Pauli matrices given
by (A.14). According to (A.12) these satisfy
$$
{\buildrel (+) \over \sigma}{\wedge}{\buildrel (+) \over \sigma} =
{\buildrel (-) \over \sigma}{\wedge}{\buildrel (-) \over \sigma} = 0,\ \ \ \ \
{\buildrel (+) \over \sigma}{\wedge}{\buildrel (-) \over \sigma} =
-{\buildrel (-) \over \sigma}{\wedge}{\buildrel (+) \over \sigma} =
2\eta.\eqno(C.14)
$$
Then one can easily prove that
$$
\gamma := \pmatrix{0 & -{\buildrel (+) \over \sigma} \cr
                   {\buildrel (-) \over \sigma} & 0 \cr}\eqno(C.15)
$$
are indeed the Dirac algebra in two dimensions. Using (A.14) one can
read off from (C.15) the explicit realization in terms of the matrices
$$
\gamma^{\hat{0}}=\pmatrix{\ 0\ & \ 1 \cr
                          -1 \ & \ 0 \cr},\ \ \ \ \
\gamma^{\hat{1}}=\pmatrix{\ 0\ & \ 1 \cr
                          \ 1\ & \ 0 \cr},\ \ \ \ \,
\gamma_5 =\pmatrix{\ 1\ & \ 0 \cr
                   \ 0 \ & -1  \cr}.\eqno(C.16)
$$

In view of (C.10) it is clear that each of the two components of the Dirac
spinor (C.1), $\psi_1$ and $\psi_2$ represent the irreducible spinor fields
with the simple transformation laws
$$
\psi^{\prime}_1 = e^{\omega / 2}\psi_1 ,\ \ \ \ \
\psi^{\prime}_2 = e^{-\omega/ 2}\psi_2 ,\eqno(C.17)
$$
compare to (A.13).

\medskip
\goodbreak
\centerline {\bf Acknowledgments}
We would like to thank R.D. Hecht for constructive discussions.
\goodbreak
\bigskip
\centerline {\bf References}
\newref
[1] C. Teitelboim, Phys. Lett. {\bf 126B} (1983) 41.
\newref
[2] R. Jackiw, Nucl. Phys. {\bf B252} (1985) 343.
\newref
[3] R. Jackiw: ``Gauge Theories for Gravity on a Line", MIT preprint
CTP\# 2105 (1993).
\newref
[4] D. Cangemi and  R. Jackiw,  Phys. Rev. Lett. {\bf 69} (1992) 233.
\newref
[5] W. Kopczy\'nski, {\it Ann. Phys.} (N.Y.) {\bf 203}, 308 (1990).
\newref
[6] E.W. Mielke, Ann. Phys. (N.Y.) {\bf 219} (1992) 78.
\newref
[7] W. Thirring: {\it Classical Field Theory} 2nd ed., Springer-Verlag,
New York/Vienna, 1980.
\newref
[8] E.W. Mielke, J.D. McCrea, Y. Ne'eman, and F.W. Hehl:
``Avoiding degenerate coframes in an affine gauge approach to quantum
gravity" (University of Cologne preprint 1992).
\newref
[9] S. Hwang and R. Marnelius, {\it Nucl. Phys.}
{\bf B271} (1986) 369.
\newref
[10] E.A. Nazarowski and Yu.N. Obukhov,  Sov. Phys.
Doklady {\bf 32} (1987) 880. [Dokl. Akad. Nauk. SSSR {\bf 297} (1987)
334].
\newref
[11] Yu.N. Obukhov and S.N. Solodukhin, Class. Quant. Grav.
 {\bf 7} (1990) 2045.
\newref
[12] L.A. Pars: {\it A Treatise on Analytical Dynamics}
(Heinemann, London 1965).
\newref
[13] F.W. Hehl and J.D. McCrea, {\it Found. Phys.} {\bf 16}, 267 (1986);
F.W.  Hehl,  "Four lectures on Poincar\'e gauge theory",
in: Proceedings of the 6th Course of the School of Cosmology and Gravitation
on {\it Spin, Torsion, Rotation,
and Supergravity}, held at Erice, Italy, May 1979, P. G. Bergmann,
V. de Sabbata, eds. (Plenum, New York 1980), p.5.
\newref
[14] E. Witten, Phys. Rev. {\bf D44} (1991) 314.
\newref
[15] C.G. Callan,Jr., S.B. Giddings, J.A. Harvey, and A. Strominger,
Phys. Rev. {\bf D45} (1992) R1005.
\newref
[16] M.O. Katanayev and I.V. Volovich, Ann. Phys. (N.Y.) {\bf 197}
(1990) 1; M.O. Katanayev, J. Math. Phys. {\bf 31} (1990) 882; {\bf 32} (1991)
2483.
\newref
[17] H. Grosse, W. Kummer, P. Pre\u snajder and D.J. Schwarz, J. Math.
Phys. {\bf 33} (1992) 382
\newref
[18] P. Baekler, E.W. Mielke, and F.W. Hehl, Nuovo Cimento {\bf 107B} (1992)
91.
\newref
[19] J.D. McCrea, Class. Quantum Gravity {\bf 9}, 553 (1992).
\newref
[20] C,J. Isham, ``Canonical quantum gravity and the problem of time",
preprint Imperial/TP/91--92/25.
\newref
[21] A.M. Polyakov, Mod. Phys. Lett. {\bf A2} (1987) 839.
\newref
[22] S.N. Solodukhin, {\it Two-dimensional black holes with torsion},
preprint JINR E2--93--33.
\newref
[23] V.P. Frolov, Phys. Rev. {\bf D46} (1992) 5383.
\newref
[24] A. Bilal and I.I. Kogan: {\it Hamiltonian Approach to 2D Dilaton--%
Gravities and Invariant ADM Mass.} hep-th/93 01 119.
\newref
[25] G.W. Gibbons and M.J. Perry: {\it The Physics of 2-d Stringy Spacetime.}
hep-th/92 04 090.
\newref
[26] C.W. Misner, K. Thorne, and J.A. Wheeler: {\it Gravitation}
(S. Freeman, San Francisco, 1973).
\newref
[27] W. Kummer and D.J. Schwarz, Nucl. Phys. {\bf B382} (1992) 171.
\newref
[28] P. Baekler, E.W. Mielke, R. Hecht, and F.W. Hehl, Nucl. Phys. {\bf
B288} (1987) 800.
\newref
[29] F.W. Hehl, J. Lemke, and E.W. Mielke, in:  Proc. of the School on
{\it Geometry and Theoretical Physics}, Bad Honnef, 12 --16 Feb. 1990,
J. Debrus and A.C. Hirshfeld, eds. (Springer, Berlin 1991), p. 56.
\newref
[30] F.W. Hehl, {\it Found. Phys.} {\bf 15}, 451  (1985).
\newref
[31] R. Hecht, F.W. Hehl, J.D. McCrea, E.W. Mielke, and Y. Ne'eman,
Phys. Lett. {\bf A172} (1992) 13.
\newref
[32] The second term
$(e_{\beta}\rfloor{\buildrel\frown\over{D}}\zeta^{\gamma})\,
\tau^{\beta}{}_{\gamma}$, contained in (10.8) vanishes quite generally.
This can be seen after substituting the Killing vector (11.1) of
the black hole solution into (10.8).
The explicit calculation is rather lengthy, but purely algebraic.  All
the necessary formulae are given in this paper, especially, in the
Appendices A and B.
\newref
[33] D. Stauffer, F.W. Hehl, V. Winkelmann, and J.G. Zabolitzky:
{\it Computer Simulation and Computer Algebra}, Third edition (Springer,
Berlin 1993).

\centerline{ ---------------------------}

\vfill\bye